\documentclass[aps,prc,twocolumn,superscriptaddress]{revtex4-1}

\usepackage{amsmath}
\usepackage{amsfonts}
\usepackage{graphicx}
\usepackage[caption=false]{subfig}
\usepackage{amssymb}
\usepackage{color}
\usepackage{bbold}
\usepackage{hyperref}
\usepackage{url}
\usepackage{tikz}
\usepackage{blkarray}
\usepackage{mhchem}

\hypersetup{
    colorlinks,
    citecolor=black,
    filecolor=black,
    linkcolor=black,
    urlcolor=black}

\newcommand{\ket}[1]{| #1 \rangle}
\newcommand{\bra}[1]{\langle #1 |}
\newcommand{\braket}[1]{\langle #1 \rangle}
\newcommand{\Ho}{\text{Ho}} 
\newcommand{\Dy}{\text{Dy}}   

\usetikzlibrary{decorations.markings}
\usetikzlibrary{decorations.pathmorphing}

\tikzset{wiggle/.style={decorate, decoration=snake}}
\tikzset{->-/.style={decoration={
  markings,
  mark=at position 0.6 with {\arrow{latex}}},postaction={decorate}}}
\tikzset{dArrow/.style={decoration={
  markings,
  mark=at position 0.7 with {\arrow{latex}}},postaction={decorate}}}

\begin{document}

\title{\textit{Ab initio} calculation of the calorimetric electron capture spectrum of $^{163}$Holmium: Intra-atomic decay into bound-states}

\author{M. Bra\ss}
\affiliation{Institute for Theoretical Physics, Heidelberg University, Philosophenweg 19, 69120 Heidelberg, Germany}

\author{C. Enss}
\affiliation{Kirchhoff Institute for Physics, Heidelberg University, Im Neuenheimer Feld 227, 61120 Heidelberg, Germany}

\author{L. Gastaldo}
\affiliation{Kirchhoff Institute for Physics, Heidelberg University, Im Neuenheimer Feld 227, 61120 Heidelberg, Germany}

\author{R. J. Green}
\affiliation{Department of Physics and Engineering Physics, University of Saskatchewan, Saskatoon, Saskatchewan, Canada S7N 5E2}
\affiliation{Stewart Blusson Quantum Matter Institute, University of British Columbia, Vancouver, British Columbia, Canada V6T 1Z4}

\author{M. W. Haverkort}
\affiliation{Institute for Theoretical Physics, Heidelberg University, Philosophenweg 19, 69120 Heidelberg, Germany}

\date{\today}

\begin{abstract}
The determination of the electron neutrino mass by electron capture in $^{163}$Ho relies on a precise understanding of the deexcitation of a core hole after an electron capture event. Here we present an \textit{ab intio} calculation of the electron capture spectrum of $^{163}$Ho, i.e. the $^{163}$Ho decay rate as a function of the energy distribution between the $^{163}$Dy daughter atom and the neutrino. Our current level of theory includes all intra-atomic decay channels and many-body interactions on a basis of fully relativistic bound-orbitals. We use theoretical methods developed and extensively used for the calculation of core level spectroscopy on correlated electron materials. Our comparison to experimental electron capture data critically tests the accuracy of these theories. We find that relativistic interactions beyond the Dirac equation lead to only minor shifts of the spectral peaks. The electronic relaxation after an electron capture event due to the modified nuclear potential leads to a mixing of different edges, but due to conservation of angular momentum of each scattered electron, no additional structures emerge. Many-body Coulomb interactions lead to the formation of multiplets and to additional peaks corresponding to multiple core-holes created via Auger decay. Multiplets crucially change the appearance of the resonances on a Rydberg energy scale. The additional structures due to Auger decay are, although clearly visible, relatively weak compared to the single core hole states and are incidentally far away from the end-point region of the spectrum. As the end-point of the spectrum is affected most by the neutrino mass, these additional states do not directly influence the statistics for determining the neutrino mass. The multiplet broadening and Auger shake-up of the main core-level edges do however change the apparent line-width and accompanying lifetime of these edges. Fitting core level edges, either in electron capture spectroscopy, or in x-ray absorption spectroscopy, by a single resonance thus leads to an underestimation of the core hole lifetime. 

\end{abstract}


\maketitle

\section{\label{intro}Introduction}
The existence of a finite neutrino mass, implied by observed neutrino flavour oscillations, is a clear indication for physics beyond the standard model of particle physics. Knowledge of the exact values of the different neutrino masses and their mixing angles can thus be used to test theories trying to extend the standard model. The experimental determination of neutrino masses, however, remains difficult as the masses are very small and neutrinos interact very weakly with other matter. In the case of electron neutrinos, the mass can be determined from the analysis of low energy electron capture or via beta decay. Presently, two nuclides are considered for the determination of the electron neutrino and anti-neutrino mass: $^{163}$Ho and $^3$H, respectively \cite{Drexlin:2013jn}. The possibility to determine the neutrino mass from the analysis of these spectra relies on a precise knowledge of the expected spectral shape for the case of massless neutrinos. 

The reason why $^{163}$Ho is the best nuclide to investigate the electron neutrino mass is that it has the smallest energy available of all possible nuclides for the electron capture process. This energy is given by the difference between the mass of the parent \ce{^{163}_{67}Ho} and daughter \ce{^{163}_{66}}Dy atoms, and corresponds to $Q_{\mathrm{EC}}\,=\,2833\,\pm \,30^{\mathrm{stat}} \, \pm \, 15^{\mathrm{syst}}$ eV \cite{Eliseev:2015cp}. This total decay energy is shared between the neutrino (kinetic energy and rest mass $E_{\nu}=\sqrt{p_{\nu}^2c^2+m_{\nu}^2c^4}$) and excitations of the resulting $^{163}$Dy atom (electronic excitations as well as the recoil energy, or excited phonons in a solid, of the $^{163}$Dy nucleus). In particular, the fact that neutrinos have a finite mass implies that the maximum energy that can be stored in the atomic excitation of the daughter atom is $Q_{\mathrm{EC}}\, - \, m_{\nu}c^2$. As a result the finite mass of electron neutrinos can be investigated by analyzing the endpoint region of the electron capture spectrum of $^{163}$Ho.

In order to enhance the sensitivity for detection of effects arising from a finite electron neutrino mass, a calorimetric measurement of the electron capture spectrum was suggested \cite{DeRujula:1982us}. The small $Q_{\mathrm{EC}}$ of $^{163}$Ho means that the fraction of events in the small energy region below the endpoint of the spectrum is large enough to allow for such a measurement. This measurement can be performed by enclosing the $^{163}$Ho source in a suitable detector able to precisely measure energies below 10 keV. In modern experiments---such as the "Electron Capture in $^{163}$Ho" experiment ECHo \cite{Gastaldo:2017ch}, "The Electron Capture Decay of $^{163}$Ho to Measure the Electron Neutrino Mass with sub-eV sensitivity" experiment HOLMES \cite{Alpert:2015gi}, and the "Neutrino Mass via Electron Capture Spectroscopy" experiment NUMECS \cite{Croce:2016dp}---small activities of $^{163}$Ho, on the order of 10 - 100 Bq, are enclosed in absorbers of low temperature microcalorimeters \cite{Enss:2005ue}. Large arrays of very high energy resolution detectors ($\Delta E_{\mathrm{FWHM}}\,<\, 3$ eV)  will be employed in these experiments in order to acquire enough events to be sensitive to deviation in the spectral shape at the end point region due to neutrinos with sub-eV mass.

In $^{163}$Ho there are 67 protons, 96 neutrons, and 67 electrons present, of which 20 electrons (the electrons in the $n$s and $n$p$_{1/2}$ shells) have a substantial overlap with the nucleus and thus directly contribute to the electron capture amplitude. This gives rise to 7 resonances labelled $M_1$ to $P_1$ and $M_2$ to $O_2$ for capture events from the 3s to 6s and 3p$_{1/2}$ to 5p$_{1/2}$ shells, respectively. The $K$ and $L$ shells are outside the spectrum window given by the small energy difference of the $^{163}$Ho and $^{163}$Dy ground-states ($Q_{\mathrm{EC}}$).

Electrons in an atom are not independent identities and, due to strong Coulomb forces, all electrons react when one electron is captured. The calorimetrically measured spectrum is thus not given by just 7 peaks, as several additional shake-up and shake-off structures, or multiplets, appear. Previous theoretical calculations of the electron capture spectrum of $^{163}$Ho stressed the importance of additional satellites that appear in these spectra
\cite{Faessler:2015dg,Faessler:2015ck,Robertson:2015dg,DeRujula:2016cp,Faessler:2017hq,Gastaldo:2017ch}. From these papers it becomes clear that a more complete understanding of the electronic relaxation after electron capture is needed in order to better describe the present experimental spectra and to reduce systematic uncertainties related to an inadequate understanding of the $^{163}$Ho spectrum close to the endpoint. This aspect is of fundamental importance to reach sub-eV sensitivity on the electron neutrino mass in $^{163}$Ho-based experiments.

In this work, we provide an important step in the quantitative understanding of the $^{163}$Ho spectrum through \textit{ab initio} calculations of the electron capture spectra restricted to the sharp resonances corresponding to bound states of a $^{163}$Ho ($^{163}$Dy) atom embedded in Au. The approach used in this paper is based on the theory of core level spectroscopy which has been extensively developed in the field of condensed matter physics \cite{deGroot:2008wo, Tanaka:1995tl, Rehr:2009eu, Haverkort:2012du, Haverkort:2014hq}, and can be extended to the calculation of electron capture \cite{Bergmann:1999dx, Glatzel:2001ia}. Core level spectroscopy is widely used to determine valuable information on the low energy states in a multitude of materials. In our present approach, methods developed in quantum chemistry (i.e. configuration interaction) are used to calculate the many-body ground-state, and Green's function methods are used to describe the electron capture process. The theoretical description we develop yields Green's function propagators describing the time evolution of a Dy atom having a multi-configurational electronic many-body wavefunction corresponding to the ground-state of Ho with one additional core hole. The result is an electron capture spectrum (up to an overall intensity scaling) restricted to bound states calculated from first principles.

The $^{163}$Ho spectral shape we have obtained with our approach agrees well with the available data and predicts additional features which could be observed once spectra with higher statistics and better energy resolution are available. In order to facilitate a detailed understanding of the electron capture resonances and of the most important processes determining the spectral line-shape, we systematically investigate the influence of different interactions. In section II we first present our final results and compare them to experimental data. Next, in subsections II.A to II.D, we subsequently examine the influence of different relaxation channels. We start with no relaxation (A) and then add relaxation due to the modified nuclear and valence potentials (B), relaxation due to inter-core-level Coulomb scattering (C), and finally relaxation that changes the occupation of the $4f$ valence shell (D). 

In the Appendix  we provide additional details on the methods used. Sections (A) and (B) focus on the ground-state, with section (A) describing the one particle orbitals and section (B) detailing the many-electron ground-state, including quantum fluctuations. Sections (C) to (E) focus on the capture process with section (C) discussing the relation between Fermi's golden rule and the Green's function propagator, section (D) describing the decoupling of nuclear, electronic, and neutrino degrees of freedom, and section (E) providing the numerical values of the capture probabilities of the different atomic orbitals. Section (F) discusses the numerical stability of our procedure, which becomes an issue due to the large difference in interactions present in our Hamiltonian. Sections (G) and (H) relate to specific effects in the electron capture spectra. Section (G) provides additional information on the mixing of principle quantum numbers due to the modified nuclear potential, as discussed in section II.B of the main text, while section (H) discusses the consequences of relativistic effects beyond the Dirac equation.

\section{\label{res}The electron capture Spectrum}
The Hamiltonian describing the electron capture process needs to include the Coulomb interaction as well as the weak nuclear force. The former describes the interactions between the electrons and the potential of the nucleus, while the latter describes the reaction of a nuclear proton and a captured (inner) shell electron to a bound neutron and free neutrino. In order to calculate the electron capture spectrum one can treat the weak interaction as a (time dependent) perturbation by defining a transition operator $T$. The operator $T$ removes an electron from the $^{163}$Ho atom and transforms a nuclear proton to neutron, while simultaneously creating a neutrino. The electron capture spectrum is then defined by Fermi's golden rule:
\begin{equation}
\frac{\mathrm{d}\Gamma}{\mathrm{d}\omega}\propto\sum_{\Psi_{\text{Dy}^*+\nu}}\left|\langle \Psi_{\text{Dy}^*+\nu} | T | \Psi_{\text{Ho}} \rangle\right|^2 \delta(E_{\text{Ho}},E_{\text{Dy}^*}+E_{\nu}),
\end{equation}
where $\Psi_{\text{Ho}}$ is the many-body ground state of a $^{163}$Ho atom including both the electrons and the nucleus, $\Psi_{\text{Dy}^*+\nu}$ is one of the many excited states of a Dy atom combined with one additional electron neutrino, and $E_{\text{Ho}}$ and $E_{\text{Dy}^*} + E_{\nu}$ are the respective energies of these states.

As the interaction between matter and neutrinos is weak, one can write the wave function of an excited Dy atom and one neutrino as a product state $\Psi_{\text{Dy}^*+\nu}=\Phi_{Z=66} \times \psi_{\Dy^*}^{e^-} \times \phi_{\nu}$, where $\Phi_{Z=66}$ is the nuclear wave-function, $\psi_{\Dy^*}^{e^-}$ is one of the electronic wavefunctions and $\phi_{\nu}$ is one of the neutrino wavefunctions.  A similar expansion can be made for the transition operator, (see Appendix  \ref{sec:Decoupling} for more information). The separation allows one to sum explicitly over all neutrino momentum states, which due to the energy-momentum relation ($E_{\nu_e}=\sqrt{p_{\nu_e}^2c^2+m_{\nu_e}^2c^4}$) creates additional coefficients in the expression for the spectral intensity \cite{DeRujula:1982us}:
\begin{eqnarray}
\frac{\mathrm{d}\Gamma}{\mathrm{d}\omega}&\propto&\sum_{\psi_{\text{Dy}^*}^{e^-}}\left|\braket{\psi_{\Dy^*}^{e^-}|T_{e^-}|\psi_{\Ho}^{e^-}}\right|^2 \\
&\times&\delta(\omega-E_{\Dy^*}+E_{\Dy})(Q-\omega)\sqrt{(Q-\omega)^2-m_{\nu}^2}\nonumber
\end{eqnarray}
where we have introduced the energy difference between the $^{163}$Ho and Dy atomic ground states as $Q=E_{\text{Ho}}-E_{\text{Dy}}$ and the energy of the excited Dy atom as $\omega=Q-E_{\nu}$.

Fermi's golden rule requires one to sum over all possible excited states of the Dy atom. There are in principle infinitely many excited states and for core level resonances  infinitely many of them each carry an infinitesimally small spectral weight \cite{Anderson:1967tr}. Therefore, summing all final states is not a practical way to calculate the spectrum. A numerically more convenient way to treat the problem of describing these spectra is to return to the Green's function formalism and time dependent perturbation theory from which Fermi's golden rule is derived. 

Replacing the delta function by a response function of a classical Harmonic oscillator and rewriting the final state energy combined with the sum over all final states as the Hamiltonian yields the Lehmann or spectral representation of the Green's function (see Appendix  \ref{sec:FermisRule} for more information):
\begin{eqnarray}
\label{eq:Spectrum}
\frac{\mathrm{d}\Gamma}{\mathrm{d}\omega}&\propto&\left(Q-\omega\right)\sqrt{\left(Q-\omega\right)^2-m_{\nu}^2} \\
\nonumber &\times&\mathrm{Im}\Big{[}\braket{\psi_{\text{Ho}}^{e^-}|T^{\dagger}_{e^-}\frac{1}{\omega + i \frac{\gamma}{2}-H_{\text{Dy}}+E_{\text{Ho}}}T_{e^-}|\psi_{\text{Ho}}^{e^-}}\hspace{0.3cm}\\
\nonumber &&\quad-\braket{\psi_{\text{Ho}}^{e^-}|T^{\dagger}_{e^-}\frac{1}{\omega + i \frac{\gamma}{2}+H_{\text{Dy}}-E_{\text{Ho}}}T_{e^-}|\psi_{\text{Ho}}^{e^-}}\Big{]},
\end{eqnarray}
where $H_{\text{Dy}}$ is the Hamiltonian describing the interaction between the electrons in the nuclear potential of the Dy atom.

The Green's function in the Lehmann representation is related to the time evolution of the state created after an electron capture event through a Fourier transform:
\begin{eqnarray}
\label{eqFT}
\frac{\mathrm{d}\Gamma}{\mathrm{d}\omega}&\propto & (Q-\omega)\sqrt{(Q-\omega)^2-m_{\nu_e}^2} \\
\nonumber &\times& \mathrm{Re}\int_0^{\infty} e^{i\omega t}\braket{\psi_{\text{Ho}}^{e^-}|T_{e^{-}}^{\dagger}(t)T_{e^{-}}(0) \\
\nonumber && \quad\quad\quad\quad\quad\quad\quad - T_{e^{-}}^{\dagger}(0)T_{e^{-}}(t)|\psi_{\text{Ho}}^{e^-}}\mathrm{d}t ,
\end{eqnarray}
with $t$ representing the time. The expectation value $\braket{\psi_{\text{Ho}}^{e^-}|T^{\dagger}_{e^-}(t)T_{e^-}(0)|\psi_{\text{Ho}}^{e^-}}$ describes the process where one starts with the wave function of an $^{163}$Ho atom in its ground state. At time $t=0$, an electron is removed from the atom by the operator $T_{e^-}$ and at the same time a proton is transformed into a neutron, changing the nuclear charge by 1. The subsequent wave function is not an eigenstate of the modified Hamiltonian and this state is allowed to time propagate up to time $t$. At time $t$, the operator $T^{\dag}_{e^-}$ recreates an electron and simultaneously transforms a neutron into a proton. Measured is the probability amplitude to return to the ground state of the $^{163}$Ho atom.

The challenge to tackle is thus to find an accurate description of the atomic $^{163}$Ho ground state as well as the time evolution of this wave function after an electron is removed and the nuclear potential changes. The many-body ground state of a $^{163}$Ho atom is approximately given by a state where one fills all orbitals of the 1s to 6s shells, 2p to 5p shells, 3d to 4d shells, and has an additional 11 electrons in the 4f shell, with local quantum numbers $L=6$, $S=3/2$, and $J=15/2$. The reason this is only the approximate ground state is twofold. First, the Coulomb repulsion is not infinitely larger than spin-orbit coupling, making an $L-S$ coupling scheme only approximately valid. Second, Coulomb scattering of electrons from filled shells into unoccupied shells mixes in other configurations. The full ground-state, and thorough details of the calculations, are discussed in more detail in Appendix  \ref{ground state}.  

\begin{figure}[t]
  \includegraphics[width=\columnwidth]{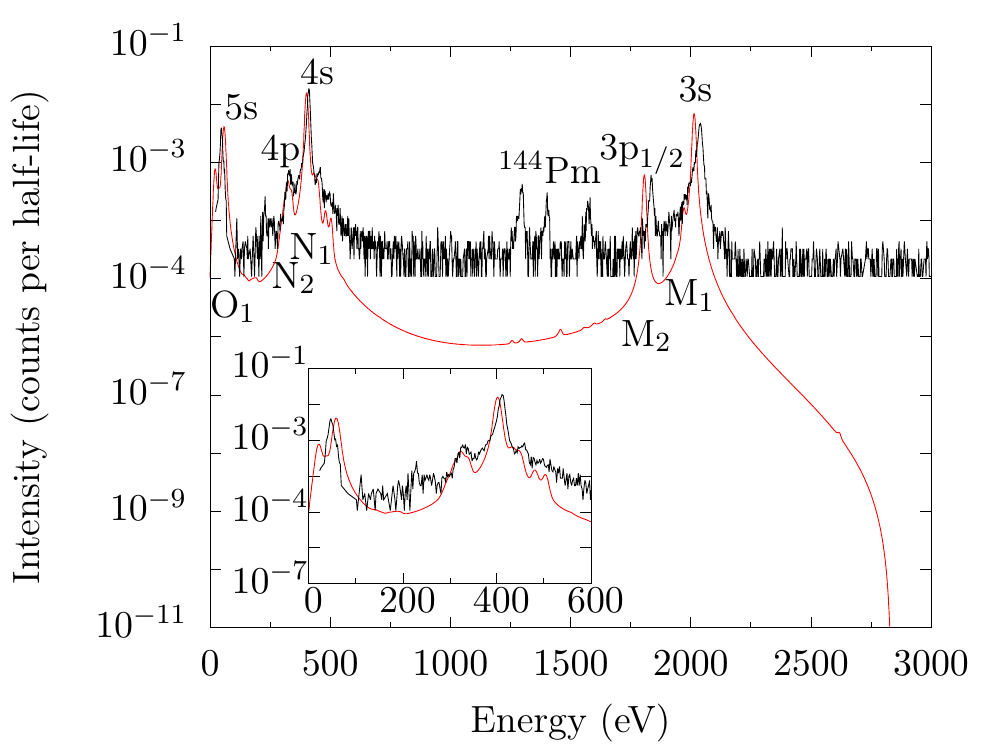}
\caption{\label{figTheoEx}Theoretical electron capture spectrum (red) assuming a constant Lorentzian line width of $5$eV, convoluted with a Gaussian distribution of $10$ eV (full width half maximum), compared to the measured spectrum (black) \cite{Ranitzsch:2017cp}. The experimental and theoretical intensities are scaled to represent the capture probability per atom, per half-life of $^{163}$Ho. Differences in the apparent line-widths of the different edges is due to the decay channels and unresolved multiplet structure underlying the edges, included in the calculation. The inset shows a shoulder and two additional peaks next to the 4s peak. These features result from Auger decay creating double vacancies in the 4p and 4d shell and an additional electron in the 4f shell. A similar shoulder is left of the 3s peak which is too small for the experimental resolution to resolve. (See Fig. \ref{figLabeled} for a high resolution theoretical spectrum with more extended labelling.)}
\end{figure}

As our calculation is restricted to bound states only, the spectrum is in principle given by a discrete set of delta functions. In order to plot the spectra and to compare them to experiment we added an additional broadening. In most of the calculations we included a Lorentzian lifetime broadening of the core hole of 1 eV full width half maximum. In Fig. \ref{figTheoEx} we compare the calculated spectrum to data obtained in calorimetric measurements. In order to find a good comparison in both peak maximum and overall line width we used an edge independent Lorentzian line width of 5 eV and convoluted the spectrum with a Gaussian distribution of 10 eV FWHM to account for detector broadening.

Given the level of theory used we find a satisfactory agreement between theory and experiment, including correct energies and relative intensities of the $M_1$, $M_2$, $N_1$, and $N_2$ edges. The shoulder structure at the high energy side of the $N_1$ edge is reproduced in our calculations with roughly the correct position and intensity of these additional structures. The maximum discrepancy between the measured and calculated peak positions is about 20 eV. We expect that this can be improved by expanding our one particle basis with states from higher shells, by including the chemical shift induced by gold surrounding the $^{163}$Ho, and by adding the self-energies of the excited states due to decay into continuum states. The self-energies are complex, with the real and imaginary parts having different effects on the spectra. While the imaginary part determines the linewidth of the edges, the real part shifts the edge energy by an amount of roughly the same order of magnitude as the imaginary part. Consequently, we expect that these corrections together would yield the experimentally observed energies.

Between the $N_1$ and $M_2$ edges, there is a discrepancy between the experimental and theoretical intensity. The shape of the tails of the resonances is not captured completely on the current level of theory. This indicates that approximating the spectrum by Lorentzian-shaped resonances of bound states is not sufficient to describe the tails of the spectrum. Explicit lifetime broadening due to Auger and fluorescence decay into continuum states should be included in future calculations. 

Additionally, we find that although all states are broadened with the same lifetime, the different resonances appear to have different widths, in agreement with experiment. This is a direct result of the decay of the core excited states due to an electron capture event into bound states, which is explicitly included in our calculations. In order to better understand these effects, in the following subsections, we continue by building the spectrum in a step by step fashion. In subsection A to D we do not include any experimental (Gaussian) broadening in the theory. Furthermore we reduced the additional broadening added in our theory to account for decay into continuum states from 5 eV to 1 eV. This leads to sharper peaks in the calculation than one would expect in the experiment, but it also allows one to discuss clearer the multiplet effects and the broadening that arrises from them.

\subsection{\label{res:no_relaxation}The electron capture spectrum without atomic relaxation}

\begin{figure}[t]
\includegraphics[width=\columnwidth]{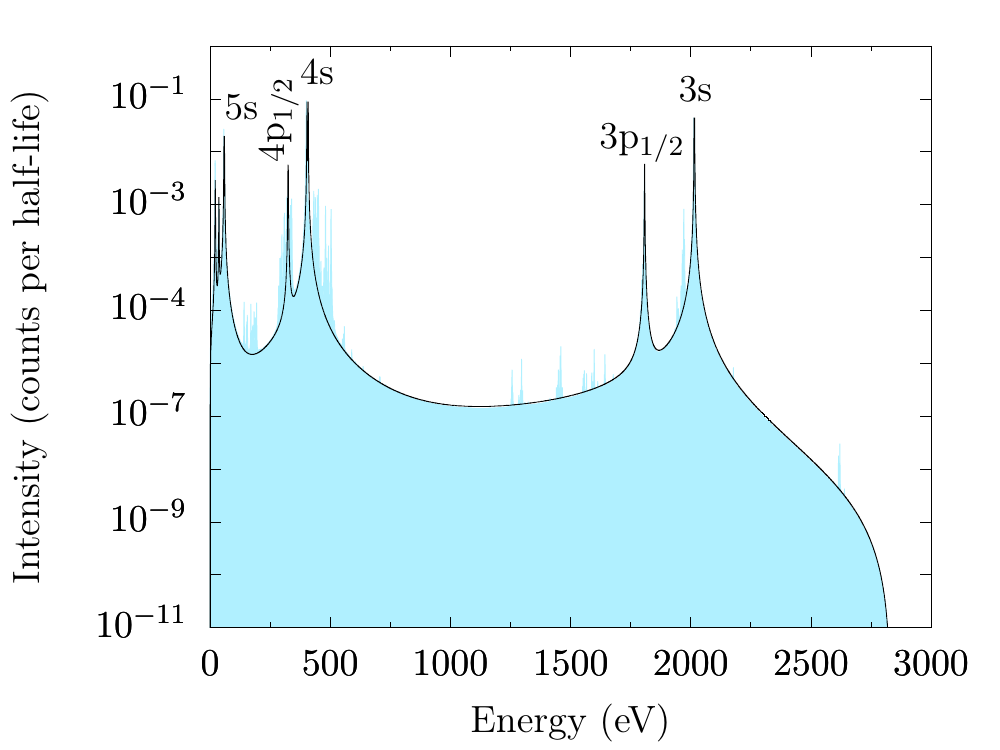}
\caption{\label{figDiagA2}The theoretical electron capture spectrum neglecting all electronic relaxation (black) compared to the full calculation (blue). The black spectrum resembles the situation shortly after the electron capture event as described in subsection \ref{res:no_relaxation}.  Directly after the capture event only the main lines (no shake-up)|  mbles the situation shortly after the electron capture event as described in subsection \ref{res:no_relaxation}. Both calculations assume a constant Lorentzian line width of $1$ apear. Hence the black spectrum does not contain the additional excitations or multiplets as in the full calculation (blue). Both calculations assume a constant Lorentzian line width of $1$eV.}
\end{figure}

At time $t=0$, the operator $T$ creates a core hole in any one of the $n$s or $n$p$_{1/2}$ orbitals of the $^{163}$Ho atom. If one would freeze the wave function into this state (i.e. $\Psi(t)=T \Psi_{\text{Ho}}$), then the spectrum would consist of separated delta functions corresponding to the $n$s and $n$p$_{1/2}$ orbitals from which the electron is captured into the nucleus. In Fig. \ref{figDiagA2} we show this spectrum (where all relaxation processes are neglected) in black. For comparison, the blue spectrum in the background shows the calculation after including all interactions and relaxations into bound orbitals.

The spectrum obtained without the inclusion of relaxation processess is already in quite good agreement with the full calculation. The full calculation does have several additional shoulders and peaks which would change peak widths at the resolution level of Fig. \ref{figTheoEx}, but the overall intensity and energy are quite reasonable for the simplified calculation. The energy shifts of the largest peaks are of the order of several electron volt and the intensity of the peaks is changed by no more then a few percent and recovered if one integrates the whole spectrum. This observation is related to sum-rules stating that further decay of the state created by electron capture can shift spectral weight, but does not change the integrated spectral weight. 

\subsection{\label{res:potential_relaxation}The electron capture spectrum including relaxation due to modified nuclear and core hole potentials}
\begin{figure}[t]
\includegraphics[width=\columnwidth]{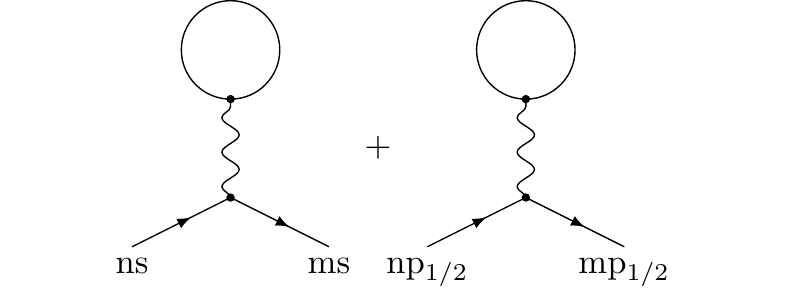}
\caption{\label{diag:PotentialRelaxation}Core hole scattering due to a change in the spherical nuclear and core hole potential after an electron capture event. These processes only allow for changes in principle quantum numbers of the core hole. Hence, the major deexcitation energies are slightly shifted, but no additional excitations emerge (see text).}
\end{figure}
The first additional relaxation process one can consider is due to the modified nuclear and core hole potentials. Since the potential of the nucleus is spherical and angular momentum is conserved, scattering of the holes is restricted to orbitals of the same angular momentum $(\kappa)$, but with different principle quantum number. In a diagrammatic language this means including the diagrams in Fig. \ref{diag:PotentialRelaxation} into the calculation of the spectrum. These diagrams only allow the created hole to scatter between states which already could have been created by the transition operator $T$ acting on the $^{163}$Ho ground state. The Hilbert space needed to describe $T_{e^{-}} \psi_{\text{Ho}}^{e^-}$ is sufficient to describe scattering due to the modified nuclear and core hole potentials. In other words, this modified interaction leads to mixing between the hole-orbitals, and induces a level repulsion between them. Due to conservation of angular momentum this spherical potentials does not lead to additional shake-up peaks. In terms of a time dependent picture, the operator $T$ can annihilate an electron from the $n$s orbital, the resulting hole then scatters into the $m$s orbital, and after a time $t$ the operator $T^{\dag}$ places the electron back. Capture events from, for example, the $^{163}$Ho $1s$ orbital thus have a significance, as these orbitals are not orthogonal to the Dy $ns$ orbitals. For these off diagonal terms it is important to remember that these scattering events can induce a change in sign and thus the corresponding contributions to the Green's function come with a phase such that the holes moving via different paths interfere with each other. These phases can change if one is above or below a resonance, leading to Fano-like lineshapes. The full Green's function matrix showing how electrons captured in the $ns$ shell can propagate to the $ms$ shell and thereby influence the electron capture spectrum is shown in Appendix  \ref{sec:mixing}.

\begin{figure}[t]
\includegraphics[width=\columnwidth]{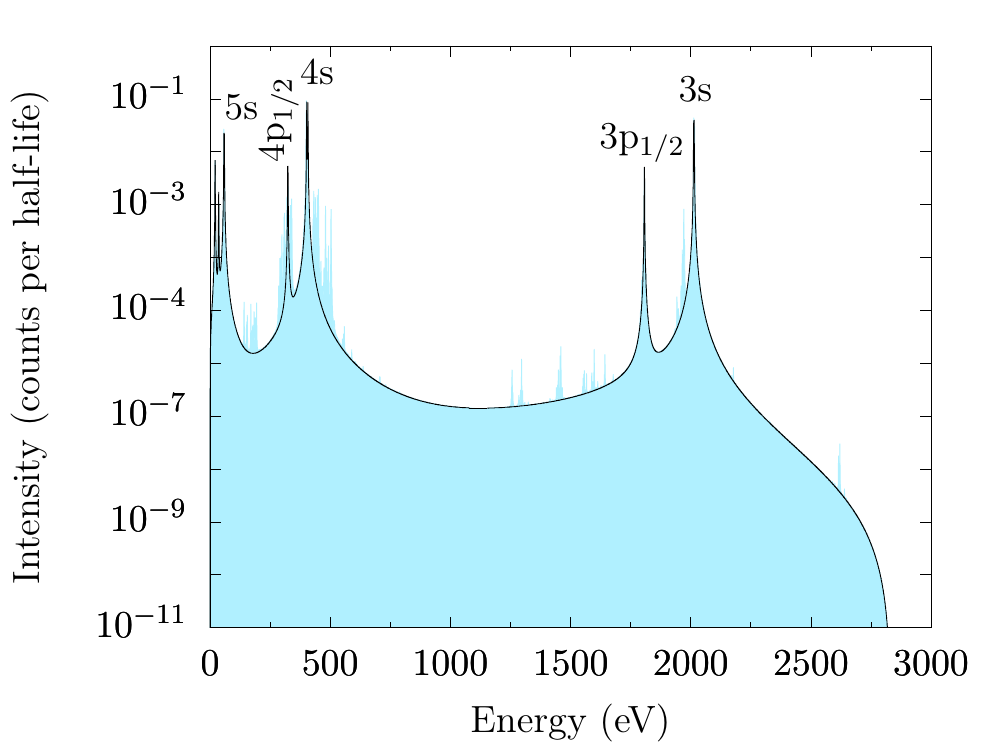}
\caption{\label{figNTri1}Theoretical electron capture spectrum including relaxation due to modified nuclear and core hole potentials (black) compared to the full calculation (blue). The black spectrum includes electron scattering which conserves the angular momenta of the scattered electron in the spherical atomic potential, as described in section \ref{res:potential_relaxation}. The peak energies shift by a maximum of $\sim3$ eV compared to the spectra in Fig.\ref{figDiagA2}.}
\end{figure}

Overall the influence of these off-diagonal elements in $^{163}$Ho leads to a shift of the major peaks of up to 3 eV, which is relatively modest. At this level of theory the major peaks are shifted to their final positions on the electron volt scale. This implies that the major excitation peaks can be understood by holes moving in a potential induced by core and valence electrons and a mixing of these holes due to the modified nuclear charge and Coulomb repulsion. For comparison we included  the spectra calculated including all scattering processes as a blue background in Fig. \ref{figNTri1}. Relaxation and mixing due to off diagonal elements do not cause the shoulders and further excitation peaks present in the full calculations. 

\subsection{\label{res:core_relaxation}The electron capture spectrum including inter-core relaxation due to Coulomb repulsion}
\begin{figure}[t]
\includegraphics[width=\columnwidth]{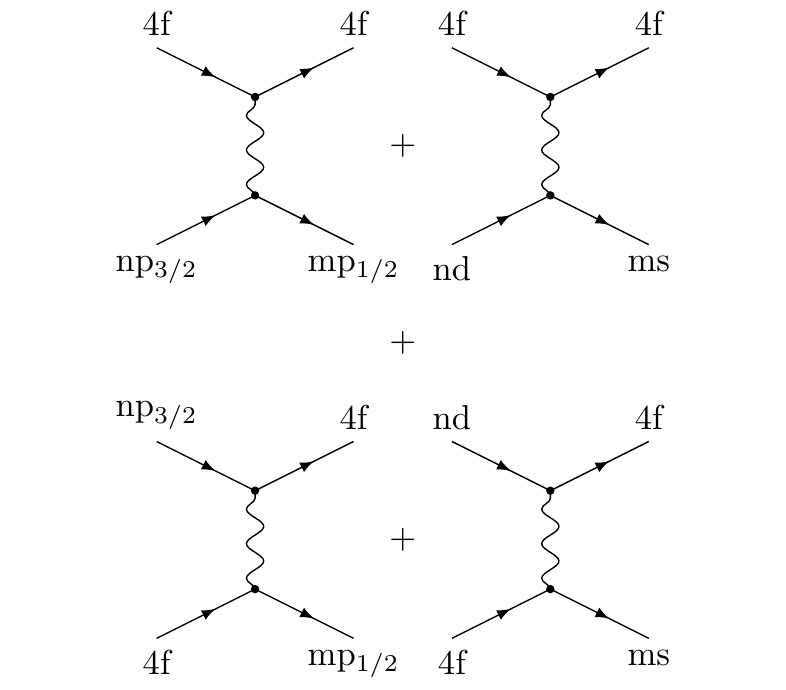}
\caption{\label{diag:InterCoreRelaxation}Coulomb repulsion between the core and $4f$ valence electrons can transfer angular momentum from the core hole to the $4f$ valence shell. These relaxation processes lead to core holes in shells from which EC is not directly possible ($p_{3/2}$ or $d$). Hence, additional spectral features emerge (see figure \ref{figRest11}). Top: direct terms. Bottom: exchange terms.}
\end{figure}

\begin{figure}[t]
  \includegraphics[width=\columnwidth]{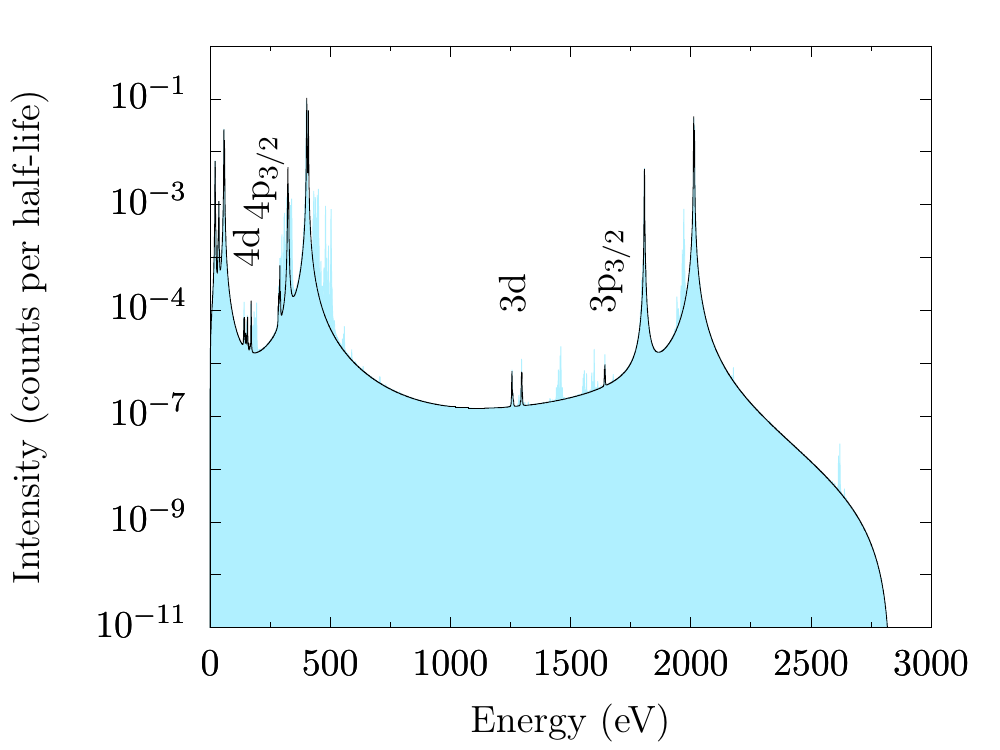}
\caption{\label{figRest11}Theoretical electron capture spectra including inter-core relaxation due to Coulomb repulsion (black) compared to the full calculation (blue). Additional resonances compared to Fig. \ref{figNTri1} appear due to core-holes in the $3p_{3/2}$, $3d_{3/2}$, and $3d_{5/2}$ orbitals in the range of 1600 to 1200 eV and core holes in the $4p_{3/2}$, $4d_{3/2}$, and $4d_{5/2}$ orbitals in the range of 400 to 200 eV.}
\end{figure}

The next level of relaxation includes scattering between core shells of different angular momentum. In a many-body calculation, the angular momentum of a single electron does not need to be conserved. Only the angular momentum of all electrons together is conserved. In Ho the $4f$ shell is partially filled and $^{163}$Ho has besides a nuclear moment a local electronic magnetic (spin and angular) moment. For the electrons one can change the angular momentum of the core hole if one simultaneously changes the alignment of this moment with respect to the valence moment. This allows for $p_{3/2}$ electrons to scatter into $p_{1/2}$ holes or $d$ electrons to scatter into $s$ holes. These interactions are given by the diagrams in Fig \ref{diag:InterCoreRelaxation}. The diagrams describe the process where a p$_{3/2}$ (d) electron scatters into a p$_{1/2}$ (s) hole (created when the electron was captured into the nucleus), transferring its angular momentum to an f electron in the valence shell. This gives rise to additional peaks shown in Fig. \ref{figRest11} which are at the excitation energies of the 3p$_{3/2}$, 3d, 4p$_{3/2}$ and 4d orbitals. The peaks are split into multiplets, as there are several ways one can achieve the alignment of the core and valence spin and angular momentum while fulfilling the conservation rules imposed on them. The probability for this process, by which the hole moves to a different excitation energy level, depends on the corresponding scattering amplitude given by the Coulomb interaction and the energy difference between the states that participate in the scattering process. Consequently, the emerging peaks are much smaller than the main edges. 

\subsection{\label{res:relaxation_all}The electron capture spectrum including relaxation into all locally bound states}

\begin{figure}[t]
\includegraphics[width=\columnwidth]{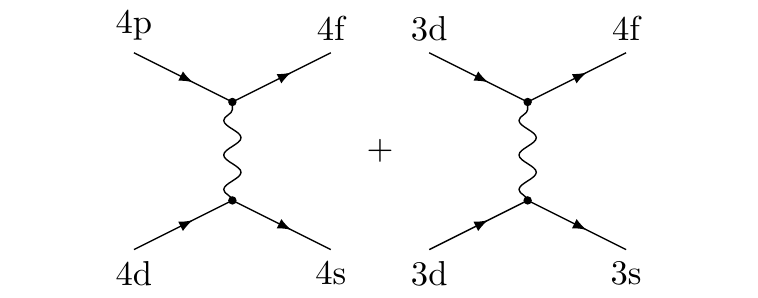}
\caption{\label{diag:TwoCoreHoles}Two of the 144 diagrams describing Auger decay including the $4f$ valence shell. After the EC event shallow core electrons can de-excite by filling the created core hole (4s or 3s in this example) and transfering energy to another shallow core electron. The later shallow core electron is transfered to the valence shell (4f) such that the atom is left with two core holes. These processes yield the additional excitations in Fig. \ref{figLabeled}. }
\end{figure}

\begin{figure}[t]
  \includegraphics[width=\columnwidth]{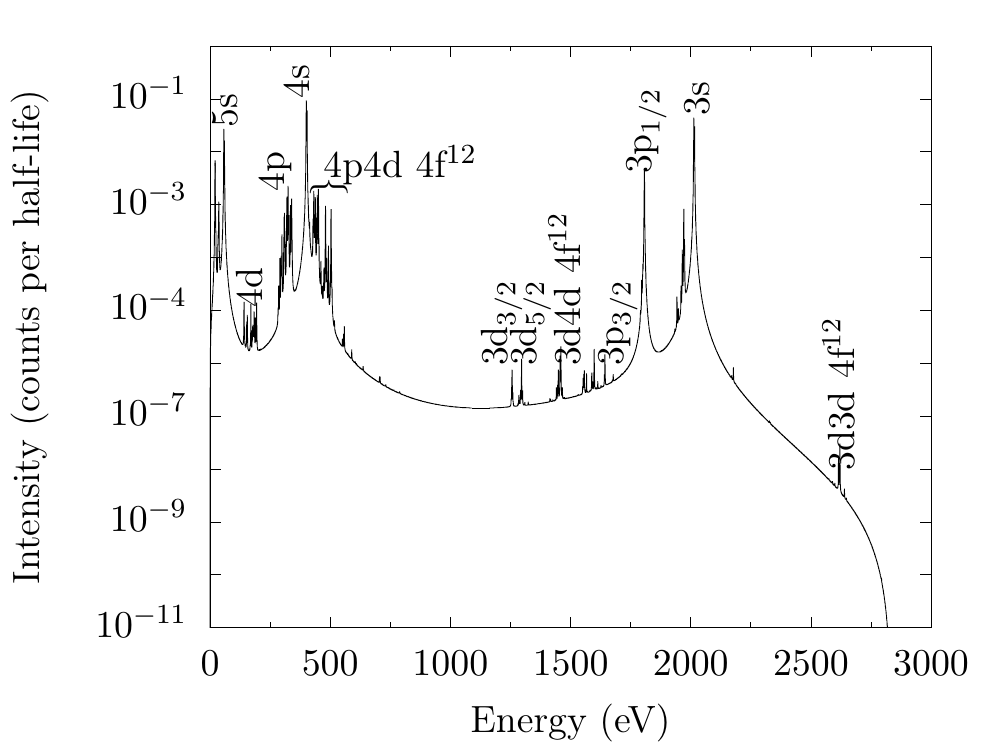}
\caption{\label{figLabeled}Theoretical spectrum including relaxation into all locally bound states, calculated with a Lorentzian linewidth of $\gamma=1$eV to reveal the contributions of the different excited states. The labels denote which shells host holes and indicate when an additional electron is present in the 4f-shell. Due to Coulomb repulsion, the peaks are split into multiplets. This spectrum is (apart from a different broadening) the same as that shown in Fig. \ref{figTheoEx}.}
\end{figure}

The last relaxation channel we include changes the number of core holes and valence electrons. Coulomb interaction allows for core electrons to scatter into the valence shell while simultaneously another core electron scatters into the previously created core hole by an electron capture event. Such processes can occur if one includes the four-point vertex where four different shells are involved in the scattering as in Fig. \ref{diag:TwoCoreHoles} for instance. However, many other diagrams are allowed. For the created electrons, one must be in the 4f shell and the other in one of the $n$s or $n$p$_{1/2}$ shells. The annihilated electrons can come from any of the occupied shells, i.e. the 1s to 6s, 2p to 5p or 3d to 4d shells. The only restriction on the scattering is that the parity of the state needs to be conserved. This results in 144 different ways to create states with two core holes, with many states having energies in the allowed range. Nonetheless, only a few strong Auger states are observed, which are labelled in Fig. \ref{figLabeled} as 4p4d4f$^{12}$, 3d4d4f$^{12}$ and 3d3d4f$^{12}$, with one core hole in each of the listed shells and 12 electrons in the 4f shell.

The corresponding scattering amplitude (i.e. the Coulomb repulsion) is large if the involved orbitals have large overlap with each other. This is the case if states have the same principle quantum number. The 4p4d4f$^{12}$ state originates from the electron capture of a 4s electron and subsequent scattering of a 4d electron into the 4s shell and a 4p electron into the 4f shell (or 4d to 4f and 4p to 4s). The 3d4d4$f^{12}$ state arises from the scattering of a 3d electron into the 3p$_{1/2}$ shell after an electron capture from this shell, and a simultaneous scattering of a 4d electron into the 4f shell. The 3d3d4f$^{12}$ state is weaker as it involves the change of principle quantum number of one of the participating electrons. A 3d electron scatters into the 3p$_{1/2}$ shell from which the electron was captured, and at the same time another 3d electron scatters into the 4f shell. 

At this point we are at a level of theory where we can understand the shoulders of the 4s and 3s peaks in Fig. \ref{figTheoEx}. These features emerge from additional excitations due to Auger decays, and are then smeared out by multiplet splitting of the double and triple open shell states involved. Experiments with higher resolution will be able to resolve these multiplets. As the intensity of possible multiplets is governed by strict selection rules that involve the valence electrons, these multiplets will strongly depend on the local symmetry of the 4f valence shell. Like x-ray absorption or core level photoemission, these line-shapes can be used to determine the valence, crystal-field state, or hybridisation strength and corresponding mixed valence of the ground state wave function of the 4f shell of $^{163}$Ho.

Lastly, we note that additional Auger peaks appear in the region between the 4s and 3p$_{1/2}$ peaks, as well as on the left flanks of 3s and 3p$_{1/2}$ peaks, which can be best seen in Fig. \ref{figLabeled}. With increased statistics these should become visible, crucially checking the validity of our approach. 

\section{\label{disc}Discussion}

An important question to answer is to what extent do these multi-core hole states influence the spectral line shape at the end-point, i.e. near $\omega=Q=2833$ eV? Here it is important to note that although there are many two-core-hole states, there is no strong state near the end-point region. The closest state visible in our calculations is the state with two holes in the 3d shell, but the intensity of this state is several orders of magnitude smaller than the intensity of the 3s state, which still dominates the spectral end-point. The fact that the spectral end-point region is dominated by single core excited states can be seen in Fig. \ref{figRest11} where the intensity of the reduced calculation at the end-point overlaps the full calculation (in blue).

There is an additional important consequence of our calculations that needs to be considered. The two-core-hole states and multiplet splitting might not directly influence the spectral end-point, but they do change the line shape of the resonances. The $N_2$ edge (core hole in the 4p shell) appears much broader than the $M_2$ edge (core hole in the 3p shell). This is a result of the larger multipole Coulomb interaction between electrons in the $4p$ and $4f$ shell compared to the interaction between electrons in the $3p$ and $4f$ shell. In general the Coulomb interaction is strongest between electrons with the same principle quantum number as these overlap more. It is this interaction between the core electrons and the open $4f$ valence shell which is largely responsible for the multiplet splitting. These multiplets effectively broaden the state at the resonance, but they do not change the lifetime of the core hole. If one does not resolve all multiplets one thus finds a peak with an apparent width that is different at resonance than in the wings. Thus, lifetimes determined experimentally close to the resonance can not be used as valid lifetimes further away. It therefore becomes crucial to include explicit calculations of the core hole lifetime (due to Auger and fluorescence decay into continuum states) in order to determine the exact shape of the end-point of the electron capture spectrum needed for an accurate determination of the neutrino mass.

\section{\label{concl}Conclusions}

We have demonstrated that methods extensively used for the calculation of core level spectra in solid state research, for example x-ray absorption or x-ray photoelectron spectroscopy, can be used to calculate the electron capture spectrum of $^{163}$Ho as measured in a calorimeter. Our \textit{ab-initio} results possess a level of accuracy which is sufficient to have predictive power compared to the current state of the art experimental spectra. Notably, our results explain the additional peaks found above the $N_1$ line as Auger decay of the 4s electron capture into a bound state with one extra 4f electron, one hole in a 4p orbital, and one hole in the 4d orbital. Our calculations also explain the extra line broadening of the $N_2$ line as an effect induced by an incidental degeneracy with Auger states. Both effects were recently observed by Ranitzsch \textit{et al}. \cite{Ranitzsch:2017cp}, but were not explained in their letter.

Future experimental spectra with higher statistics will show additional peaks on the low energy shoulder of the $M$ edges as well as one additional peak relatively close to the spectral end-point due to a state with two core holes in the 3d shell. Spectra with improved energy resolution will resolve several of the multiplet features revealed in our calculations. These features can be used as an internal consistency check, as the intensity distribution among the multiplets within one shell contains detailed information on the local symmetry, valence and crystal-field splittings of the $^{163}$Ho 4f ground state.

\section{Acknowledgments}

Part of this research was funded by the Deutsche Forschungsgemeinschaft (DFG, German Research Foundation) - 400329440, Research Unit FOR2202 Neutrino Mass Determination by Electron Capture in 163Ho, ECHo (funding under Grants No. GA2219/2-1 and No. EN299/7-1).

\clearpage

\appendix

\section{\label{basis}One particle spin-orbitals and many electron states}

The one particle orbitals used as a basis in our calculations are fully relativistic, numerical atomic orbitals calculated on an interpolated logarithmic grid. The spin-orbitals are formally labelled by the principle quantum number $n$ and the relativistic angular momentum quantum number $\kappa$. We adopt the notation to label the spin-orbitals by the angular momentum ($l$) of the large part of the wave function and the total angular momentum $j$, in line with the non-relativistic labelling of these orbitals.

The basis orbitals are calculated using a finite size nucleus in order to capture the overlap of the $ns$ and $np_{1/2}$ orbitals with the nucleus. This overlap defines the transition matrix elements. The final many-body calculations use the full Dirac-Coulomb-Breit interaction. In order to capture most of the change in charge density due to charge fluctuations into highly excited orbitals, the basis orbitals are calculated self-consistently on a density functional theory level using FPLO \cite{Koepernik:1999uw, Opahle:1999tx, Eschrig:2004wn}. We choose density functional theory as our starting point for the basis orbitals over Hartree-Fock orbitals as minimising the error in energy differences is more important then minimising the ground-state energy. 

Many-body states $\Psi$ can be written as linear combinations over Slater determinants $\phi$.
\begin{equation}
\Psi=\sum_i \alpha_i \phi_i,
\end{equation}
where $\alpha_i$ are numerical factors defining the state and $\sum_i |\alpha_i|^2=1$ to normalize the state. For $N$ electrons, the set of Slater determinants $\phi_i$ is given by all subsets $D_i$ of length $N$ of the possible spin-orbitals given by the principle quantum number $n$ and the angular momenta $l$, $j=l\pm1/2$ and $m$.
\begin{equation}
\ket{\phi_i} = \prod_{\{n,l,j,m\} \in D_i}c^{\dag}_{nljm}\ket{0}.
\end{equation}

\section{\label{ground state}Atomic ground state of $^{163}$Ho}

\begin{figure}[t]
\includegraphics[width=\columnwidth]{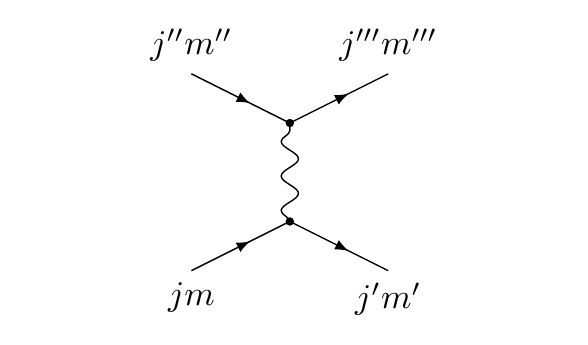}
\caption{\label{diag:JM}Coulomb scattering vertex where angular momentum $j$ is transferred. As Coulomb repulsion conserves only total angular momentum $J$, the ground state can neither be described in an $L$--$S$ coupling scheme nor a $j$--$j$ coupling scheme (see text).}
\end{figure}

The Hund's rule ground state of neutral $^{163}$Ho with configuration [Xe]6s$^{2}$4f$^{11}$ has $L=6$, $S=3/2$ and $J=15/2$. While Coulomb repulsion (and corresponding multiplet splitting) in the Lanthanide series (order of 10 eV) is much larger than spin-orbit coupling (order of 100 meV), the ground state is not a pure state in an $L$--$S$ coupling scheme. The many-body calculations are done using \textsc{Quanty}, a many-body script language developed for spectroscopy calculations \cite{Haverkort:2016hz}. For the ground state in our calculation we find $J=15/2$ ($J(J+1)=63.75$), $L=6.07$ ($L(L+1)\approx 42.90$) and $S=1.45$ ($S(S+1)\approx 3.56$). These numbers are close to $L$--$S$ coupling scheme values, but not exact. One can understand the ground state in a perturbative fashion starting from a $j$--$j$ coupling scheme ground state. In a $j$--$j$ coupling scheme, the $j=5/2$ shell is completely occupied and the remaining five electrons go into the $j=7/2$ shell. For the state with $J_z=-J$, the only unoccupied spin-orbitals would be those with $j=7/2$ and $m=3/2$, $m=5/2$ or $m=7/2$. Coulomb interaction allows scattering of the form depicted in Fig. \ref{diag:JM} with the condition that the z-component of total angular momentum is conserved, i.e. $m+m''=m'+m'''$. Thus, the Coulomb interaction allows electrons to scatter from the $j=5/2$ and $m=3/2$ or $m=5/2$ into the $j=7/2$ spin-orbital with the same $m$. Scattering into states with $j=7/2$ and $m<3/2$ is forbidden due to the Pauli principle. Additionally, scattering into the $m=7/2$ state is prohibited by conservation of angular momentum. Overall, the amount of scattering between the spin-orbitals can be best quantified by looking at the single particle density matrix.
\begin{widetext}
\begin{equation}
\label{eqDensityMatrix}
\rho_{jm,j'm'}\equiv\braket{\psi^{e^{-}}_{\text{Ho}}|c^{\dagger}_{4f_{jm}}c_{4f_{j'm'}}|\psi^{e^{-}}_{\text{Ho}}}\approx\,\,\,\begin{blockarray}{llllllllllllllr}
-\frac{5}{2}&-\frac{3}{2}&-\frac{1}{2}&\frac{1}{2}&\frac{3}{2}&\frac{5}{2}\hspace{0.75cm}&-\frac{7}{2}&-\frac{5}{2}&-\frac{3}{2}&-\frac{1}{2}&\frac{1}{2}&\frac{3}{2}&\frac{5}{2}&\frac{7}{2}&\hspace{1.0cm}m\vspace{0.25cm}\\
\begin{block}{(rrrlllrrrrllll)r}
 1 & 0 & 0 & 0 & 0 & 0 & 0 & 0 & 0 & 0 & 0 & 0 & 0 & 0 &-\frac{5}{2}\\
 0 & 1 & 0 & 0 & 0 & 0 & 0 & 0 & 0 & 0 & 0 & 0 & 0 & 0 &-\frac{3}{2}\\
 0 & 0 & 1 & 0 & 0 & 0 & 0 & 0 & 0 & 0 & 0 & 0 & 0 & 0 &-\frac{1}{2}\\
 0 & 0 & 0 & 1 & 0 & 0 & 0 & 0 & 0 & 0 & 0 & 0 & 0 & 0 &\frac{1}{2}\\
 0 & 0 & 0 & 0 & 0.88 & 0 & 0 & 0 & 0 & 0 & 0 & 0.33 & 0 & 0 &\frac{3}{2}\\
 0 & 0 & 0 & 0 & 0 & 0.94 & 0 & 0 & 0 & 0 & 0 & 0 & 0.23 & 0 &\frac{5}{2}\vspace{0.25cm}\\
 0 & 0 & 0 & 0 & 0 & 0 & 1 & 0 & 0 & 0 & 0 & 0 & 0 & 0 &-\frac{7}{2}\\
 0 & 0 & 0 & 0 & 0 & 0 & 0 & 1 & 0 & 0 & 0 & 0 & 0 & 0 &-\frac{5}{2}\\
 0 & 0 & 0 & 0 & 0 & 0 & 0 & 0 & 1 & 0 & 0 & 0 & 0 & 0 &-\frac{3}{2}\\
 0 & 0 & 0 & 0 & 0 & 0 & 0 & 0 & 0 & 1 & 0 & 0 & 0 & 0 &-\frac{1}{2}\\
 0 & 0 & 0 & 0 & 0 & 0 & 0 & 0 & 0 & 0 & 1 & 0 & 0 & 0 &\frac{1}{2}\\
 0 & 0 & 0 & 0 & 0.33 & 0 & 0 & 0 & 0 & 0 & 0 & 0.12 & 0 & 0 &\frac{3}{2}\\
 0 & 0 & 0 & 0 & 0 & 0.23 & 0 & 0 & 0 & 0 & 0 & 0 & 0.06 & 0 &\frac{5}{2}\\
 0 & 0 & 0 & 0 & 0 & 0 & 0 & 0 & 0 & 0 & 0 & 0 & 0 & 0 &\frac{7}{2}\\
\end{block}
\end{blockarray}
\end{equation}
Here $m,\,m'$ denote the z-component of angular momenta $j,\, j'$. The first six columns (rows) correspond to $j=5/2$ ($j'=5/2$) and the eight remaining ones to $j=7/2$ ($j'=7/2$). 
\end{widetext}

Scattering into the $4f_{j=7/2}$ shell does not only happen starting from the $4f_{j=5/2}$ shell, but can also happen starting from the $4d$ or even $3d$ shell, with the additional side condition that two electrons must scatter simultaneously from the $d$ to $f$ shell in order to conserve parity. Once two electrons are scattered into the 4f shell, further scattering processes into the new empty core states are possible, which influences the density matrices of the core states further.

\section{\label{sec:FermisRule}Relation between Fermi's golden rule and the Green's function propagator describing time evolution}

Most recent theoretical calculations of electron capture spectra start from Fermi's Golden Rule \cite{DeRujula:1982us,Faessler:2015dg,Faessler:2015ck,Robertson:2015dg,DeRujula:2016cp,Faessler:2017hq,Gastaldo:2017ch}. This formalism requires one to sum over all final states after an electron capture event. As the energy of the excited atom is above the auto-ionisation energy threshold, there are infinitely many of these states, each with an infinitesimally small spectral weight \cite{Anderson:1967tr}. The resulting spectrum is not given by a set of perfect Lorentzian shaped lines, but consists of multiple resonances with possible asymmetric line shapes \cite{Nozieres:1969fk,Doniach:1970wr}. These effects can be well treated using Green's functions describing the core level spectrum \cite{Haverkort:2014hq},  which is the method of choice for our calculations.

In this section we obtain the expression for the electron capture spectrum in terms of the Green's function propagator starting from Fermi's Golden Rule. In most text books these equations would be derived from time dependent perturbation theory \cite{AltlandSimons}. Here we start from Fermi's Golden Rule and from this recreate the linear response functions used in frequency and time domain. This allows us to connect to the current electron capture literature \cite{DeRujula:1982us,Faessler:2015dg,Faessler:2015ck,Robertson:2015dg,DeRujula:2016cp,Faessler:2017hq,Gastaldo:2017ch} and still end with a formalism that can be solved with diagrammatic methods. For a general transition between state $\Psi_i$ and a set of states $\Psi_f$ induced by the transition operator $T$, Fermi's Golden Rule states that the transition rate $\Gamma$ at energy $\omega$ is given by:
\begin{equation}
\label{eq:crossectionFermi}
\frac{\mathrm{d}\Gamma}{\mathrm{d}\omega}\propto\sum_f \left|\braket{\Psi_{f}|T|\Psi_{i}}\right|^2 \delta\left(E_f-\omega-E_i\right).
\end{equation}
For a diagrammatic approach to the spectrum, or an approach in terms of propagators, one has to consider virtual excitations on top of the energetically allowed excitations which are encoded in the above delta function. This is achieved by replacing the delta function by the response function of a classical damped harmonic oscillator at eigen frequency $\omega_0=E_f-E_i$ and damping $\gamma$:  
\begin{widetext}
\begin{eqnarray}
\label{eq:deltafunction}
\nonumber\delta\left(E_f-\omega-E_i\right) &\rightarrow &-\lim_{\gamma\to0^+}\mathrm{Im}\left[ \frac{E_f-E_i}{\omega^2-(E_f-E_i)^2+\mathrm{i}\gamma\omega} \right]\hspace{1cm}\\
&=&-\lim_{\gamma\to0^+}\mathrm{Im}\left[ \frac{1}{\omega-(E_f-E_i)+\mathrm{i}\gamma/2}-\frac{1}{\omega+(E_f-E_i)+\mathrm{i}\gamma/2} \right],
\end{eqnarray}

After factoring one finds two terms, one resonating at positive frequencies and one resonating at negative frequencies. This additional term at negative frequencies can be directly deduced from linear response theory \cite{AltlandSimons}, and arises naturally from the causal response to the electron capture event. In the limit of infinitesimal width the term does not contribute to the spectrum at positive frequencies $\omega$ and one thus recovers Fermi's Golden Rule as there are no measurable excitations at negative frequencies. However, in numerical calculations with finite $\gamma$, the additional term accounts for virtual excitations which have tails. These tails slightly modify the spectral shape at small positive frequencies.

Inserting the replacement of the delta function (Eq. \ref{eq:deltafunction}) in Eq. \ref{eq:crossectionFermi} and at the same time expand the square of the expectation value of the transition operator one gets
\begin{eqnarray}
\nonumber\frac{\mathrm{d}\Gamma}{\mathrm{d}\omega}\propto-\lim_{\gamma\to0^+}\mathrm{Im}\sum_f&\Big[&\braket{\Psi_{i}|T^{\dag}|\Psi_{f}} \frac{1}{\omega-(E_f-E_i)+\mathrm{i}\gamma/2}\braket{\Psi_{f}|T|\Psi_{i}}\\
&-&\braket{\Psi_{i}|T^{\dag}|\Psi_{f}} \frac{1}{\omega+(E_f-E_i)+\mathrm{i}\gamma/2}\braket{\Psi_{f}|T|\Psi_{i}}\Big].
\end{eqnarray}
As the final states $\Psi_f$ define a complete set ($\sum_f \ket{\Psi_f}\bra{\Psi_f}=\mathbb{1}$) of eigenstates of the Hamiltonian ($H \Psi_f = E_f \Psi_f$), we can replace the operator $\sum_f \ket{\Psi_f}g(E_f)\bra{\Psi_f}$ for any function $g$ by the same function acting on the Hamiltonian. Doing so yields an expression of the Green's function in the Lehmann spectral representation
\begin{equation}
\frac{\mathrm{d}\Gamma}{\mathrm{d}\omega}\propto-\lim_{\gamma\to0^+}\mathrm{Im} \Big[\left\langle\Psi_{i}\left| T^{\dag} \frac{1}{\omega-(H-E_i)+\mathrm{i}\gamma/2}T \right|\Psi_{i}\right\rangle - \left\langle\Psi_{i}\left| T^{\dag} \frac{1}{\omega+(H-E_i)+\mathrm{i}\gamma/2}T \right|\Psi_{i}\right\rangle\Big].
\end{equation}
\end{widetext}
Here we have changed the computational task of calculating all eigenstates of the Hamiltonian into the problem of finding the resolvent of the Hamiltonian evaluated for a single state. The latter can be performed using diagrammatic expansion techniques known from quantum field theory as well as Lanczos routines for finite size Hilbert spaces. For the numerical calculations we replace $\gamma$ by a small but finite width instead of taking the limit $\gamma \to 0^+$.

The relation between the spectral (or Lehmann) representation of the Green's function and the time evolution of the system becomes clear if one Fourier transforms the Green's function, which yields:
\begin{equation}
\frac{\mathrm{d}\Gamma}{\mathrm{d}\omega}\propto\mathrm{Re} \int_{0}^{\infty} e^{i\omega t}\braket{\Psi_i|T^{\dagger}(t)T(0) - T^{\dagger}(0)T(t)|\Psi_i}\mathrm{d}t
\end{equation}
where $T(t)=e^{iHt}Te^{-iHt}$ is the transition operator in the Heisenberg picture. The expectation value $\braket{\Psi_i|T^{\dagger}(t)T(0)|\Psi_i}$ describes the probability amplitude that the system excited into state $T\ket{\Psi_i}$ at time $t=0$ remains in that state after time $t$. We thus can describe the electron capture spectrum by removing an electron at $t=0$ and looking at the time evolution of the newly created state.

\section{\label{sec:Decoupling}Decoupling of electronic, nuclear and neutrino degrees of freedom}

Since electron capture involves the atomic nucleus, the electrons of the atom, as well as the created neutrino, all of these particles must be included in the full wavefunction. It is useful to decompose the full wavefunction into an electron wavefunction a nuclear wavefunction and a neutrino wavefunction whose product builds the full wavefunction. A similar decoupling can then be performed on the electron capture operator describing the transition from $^{163}$Ho to $^{163}$Dy. In this section we present how these different sectors of Fock space can be decoupled and how to construct the electron capture operator acting on each of these sectors.

The wavefunctions $\Psi$ include the electrons as well as the nucleus and possible neutrinos. The function $\Psi_{\Ho}$ represents the atomic ground state of $^{163}$Ho restricted to the sector where there are no free neutrinos available. The functions $\Psi_{\Dy^*+\nu_e}$ represent all states of the $^{163}$Dy atom, including all possible electronic excitations, plus one electron neutrino. Due to weak interaction between the sectors in the Hamiltonian containing a different number of neutrinos and modified nuclear charge, the wavefunctions can be decomposed as direct product states of a nuclear wavefunction ($\Phi_{\mathrm{Z}}$), an electron wavefunction ($\psi^{e^-}$) and a neutrino wavefunction ($\phi_{\nu}$):
\begin{eqnarray}
\ket{\Psi_{\Ho}} & = & \ket{\Phi_{Z=67}}\otimes\ket{\psi_{\Ho}^{e^-}}\otimes\ket{0}\\
\ket{\Psi_{\Dy^*+\nu}} & = & \ket{\Phi_{Z=66}}\otimes\ket{\psi_{\Dy^*}^{e^-}}\otimes\ket{\phi_{\nu}}
\end{eqnarray}
The electron capture event acts on these product states by removing a proton from $\Phi_{Z=67}$ while adding a neutron, removing an inner shell electron from $\psi_{\Ho}^{e^-}$ and creating an electron neutrino out of the vacuum $\ket{0}$. This is encoded in the electron capture operator which is given in second quantized language as 
\begin{equation}
T_{\text{tot}} = \sum_{\substack{n,l,j,m \\ k^r_{\nu},l_{\nu},j_{\nu},m_{\nu}}}c_{k^r_{\nu},l_{\nu},j_{\nu},m_{\nu}}^{n,l,j,m} T_{\text{nuclear}} T_{\text{electron}}^{n,l,j,m} T_{\text{neutrino}}^{k^r_{\nu},l_{\nu},j_{\nu},m_{\nu}}
\end{equation}
where
\begin{eqnarray}
T_{\text{electron}}^{n,l,j,m} & = & a_{\psi^{e^-},n,l,j,m}\nonumber\\
T_{\text{neutrino}}^{k^r_{\nu},l_{\nu},j_{\nu},m_{\nu}} & = & a^{\dagger}_{\phi_{\nu},k^r_{\nu},l_{\nu},j_{\nu},m_{\nu}}
\end{eqnarray}
and
\begin{equation}
T_{\text{nuclear}} \Phi_{\ce{^{163}_{67}\mathrm{Ho}}} \propto \Phi_{\ce{^{163}_{66}\mathrm{Dy}}},
\end{equation}
where we used that there is one unique ground-state in terms of total angular momentum and parity for both the $^{163}$Ho ($J=7/2$ parity odd) and $^{163}$Dy ($J=5/2$ parity odd) nuclear wave-function. Since the neutrino interacts only via weak force, we assume that the coefficients factor as $c_{k^r_{\nu},l_{\nu},j_{\nu},m_{\nu}}^{n,l,j,m}=c^{\psi_{e^-}}_{n,l,j,m}\times c_{k^r_{\nu},l_{\nu},j_{\nu},m_{\nu}}^{\phi_{\nu}}$. The neutrino part $c_{k^r_{\nu},l_{\nu},j_{\nu},m_{\nu}}^{\phi_{\nu}}$ can be neglected for $l_{\nu}\neq 0$ and is approximately constant otherwise. The full coefficient $c_{k^r_{\nu},l_{\nu},j_{\nu},m_{\nu}}^{n,l,j,m}\approx p_{nlj}$ is, up to an overall scaling constant. We approximated $p_{nlj}$ by the 
overlap between nucleus and orbital wavefunctions (See Appendix E for explicit calculations). This yields, including conservation of angular momentum, to leading order a non-vanishing contribution for $nlj\in \lbrace \text{1s - 6s, 2p$_{1/2}$ -5p$_{1/2}$} \rbrace$.

With the above decomposition of wavefunctions and electron capture operator we can now factor Fermi's Golden Rule. Starting from the asymptotic transition rate
\begin{eqnarray}
R_{\Psi_{\Ho}\rightarrow\Psi_{\Dy^*+\nu}}\left(E_{\Dy^*}+E_{\nu}\right)\propto\delta\left(E_{\Dy^*}+E_{\nu}-E_{\Ho}\right)\nonumber\\
\times\left|\braket{\Psi_{\Dy^*+\nu}|T|\Psi_{\Ho}}\right|^2\hspace{0.3cm}
\end{eqnarray}
we can express Fermi's Golden Rule as
\begin{equation}
\Gamma\propto\int\!\mathrm{d}\omega\rho (\omega)R_{\Psi_{\Ho}\rightarrow\Psi_{\Dy^*+\nu}}(\omega)
\end{equation}
where the density of states is denoted as	
\begin{equation}
\rho (\omega)\equiv\sum_{\Psi_{\Dy^*+\nu_e}}\delta\left(\omega-E_{\Dy^*}-E_{\nu}\right)
\end{equation}
\begin{widetext}
The sum runs over all exited Dy states plus a single electron neutrino $\sum_{\Psi_{\Dy^*+\nu_e}}=\sum_{\psi_{\text{Dy}^*}^{e^-}}\sum_{q^r_{\nu},l_{\nu},j_{\nu},m_{\nu}}$. Using the decomposition of the wavefunctions and the electron capture operator, the transition rate can be written as
\begin{eqnarray}
\Gamma&\propto&\int\!\mathrm{d}\omega\sum_{\psi_{\text{Dy}^*}^{e^-}}\sum_{q^r_{\nu},l_{\nu},j_{\nu},m_{\nu}}\delta\left(\omega-E_{\Dy^*}-E_{\nu}\right)\delta\left(\omega-E_{\Ho}\right)\phantom{\hspace{6cm}}\\
&\times&\left|\sum_{n,l,j,m,k^r_{\nu},l_{\nu}',j_{\nu}',m_{\nu}'}c_{k^r_{\nu},l_{\nu}',j_{\nu}',m_{\nu}'}^{n,l,j,m}\braket{\Phi_{Z=66}|T_{\text{nuclear}}|\Phi_{Z=67}}\braket{\psi_{\Dy^*}^{e^-}|T_{\text{electron}}^{n,l,j,m}|\psi_{\Ho}^{e^-}}\braket{\phi_{\nu}\left(q^r_{\nu},l_{\nu},j_{\nu},m_{\nu}\right)|T_{\text{neutrino}}^{k^r_{\nu},l_{\nu}',j_{\nu}',m_{\nu}'}|0}\right|^2.\nonumber
\end{eqnarray}
\end{widetext}
Explicitly calculating the matrix elements yields
\begin{multline}
\braket{\phi_{\nu}\left(q^r_{\nu},l_{\nu},j_{\nu},m_{\nu}\right)|T_{\text{neutrino}}^{k^r_{\nu},l_{\nu}',j_{\nu}',m_{\nu}'}|0}=\\
\braket{\phi_{\nu}\left(q^r_{\nu},l_{\nu},j_{\nu},m_{\nu}\right)|\phi_{\nu}\left(k^r_{\nu},l_{\nu}',j_{\nu}',m_{\nu}'\right)}=\\
\delta_{q^r_{\nu},k^r_{\nu}}\delta_{l_{\nu},l_{\nu}'}\delta_{j_{\nu},j_{\nu}'}\delta_{m_{\nu},m_{\nu}'}
\end{multline}
\begin{equation}
\braket{\Phi_{Z=66}|T_{\text{nuclear}}|\Phi_{Z=67}}\propto\braket{\Phi_{Z=66}|\Phi_{Z=66}}=1
\end{equation}
\begin{equation}
\sum_{n,l,j,m}c_{q^r_{\nu},l_{\nu},j_{\nu},m_{\nu}}^{n,l,j,m}\braket{\psi_{\Dy^*}^{e^-}|T_{\text{electron}}^{n,l,j,m}|\psi_{\Ho}^{e^-}}=\braket{\psi_{\Dy^*}^{e^-}|T_{e^-}|\psi_{\Ho}^{e^-}},
\end{equation}
where
\begin{equation}
T_{e^-}=\sum_{n,l,j,m}p_{n,j}T_{\text{electron}}^{n,l,j,m}
\end{equation}

We introduce a shift of variable $\omega\rightarrow\omega+E_{\Dy}+E_{\nu}$ to assure that $\omega$ represents the (calorimetrically measured) deposited energy. The electron capture spectrum is obtained by taking the derivative with respect to $\omega$:
\begin{eqnarray}
\frac{\mathrm{d}\Gamma}{\mathrm{d}\omega}\propto\sum_{\psi_{\text{Dy}^*}^{e^-}}\sum_{q^r_{\nu},l_{\nu},j_{\nu},m_{\nu}}\delta\left(\omega-E_{\Dy^*}+E_{\Dy}\right)\hspace{2cm}\\
\nonumber\times\delta\left(\omega+E_{\Dy}+E_{\nu}-E_{\Ho}\right)\left|\braket{\psi_{\Dy^*}^{e^-}|T_{e^-}|\psi_{\Ho}^{e^-}}\right|^2
\end{eqnarray}

Here the neutrinos are completely decoupled and consequently free particles such that the sum over the neutrino states can easily be evaluated as an integral over the neutrino's kinetic energy plus rest mass
\begin{equation}
\sum_{q^r_{\nu}}\propto\int_0^{\infty}\! \mathrm{d}E_{\nu}\, E_{\nu}\sqrt{E^2_{\nu}-m_{\nu}^2}
\end{equation}
Therefore, the spectral function reads
\begin{eqnarray}
\label{eq:spec1}
\frac{\mathrm{d}\Gamma}{\mathrm{d}\omega}&\propto&\sum_{\psi_{\text{Dy}^*}^{e^-}}\left|\braket{\psi_{\Dy^*}^{e^-}|T_{e^-}|\psi_{\Ho}^{e^-}}\right|^2 
\\
&\times&\delta(\omega-E_{\Dy^*}+E_{\Dy})(Q-\omega)\sqrt{(Q-\omega)^2-m_{\nu_e}^2}\nonumber
\end{eqnarray}
Now we repeat the steps from Appendix \ref{sec:FermisRule} and replace the delta distribution by a Lorentzian such that we arrive at the final result
\begin{eqnarray}
\label{eqSpectrum}
\frac{\mathrm{d}\Gamma}{\mathrm{d}\omega}&\propto &\left(Q-\omega\right)\sqrt{\left(Q-\omega\right)^2-m_{\nu}^2}\\ &\times &\mathrm{Im}\Big{[}\braket{\psi_{\text{Ho}}^{e^-}|T^{\dagger}\frac{1}{\omega + i \frac{\gamma}{2}-H_{\text{Dy}}+E_{\text{Dy}}}T|\psi_{\text{Ho}}^{e^-}}\nonumber\\
&-&\braket{\psi_{\text{Ho}}^{e^-}|T^{\dagger}\frac{1}{\omega + i \frac{\gamma}{2}+H_{\text{Dy}}-E_{\text{Dy}}}T|\psi_{\text{Ho}}^{e^-}}\Big{]}\nonumber
\end{eqnarray}

\section{\label{calc:matrix_elements}Electron capture transition matrix elements}
As described in Appendix \ref{sec:Decoupling} the electron capture operator acting on the electrons is given as
\begin{equation}
T_{e^-}=\sum_{nljm}p_{nlj} a_{nljm}\hspace{0.5cm}nlj\in \lbrace \text{1s - 6s, 2p$_{1/2}$ -5p$_{1/2}$} \rbrace.
\end{equation}
Here the quantum numbers $nlj$ label the ten shells with large cross sections in $^{163}\mathrm{Ho}$. The probability amplitude, $p_{nlj}$, is approximated to be proportional to the overlap between the electron and nuclear wavefunctions. For a nucleus of constant density and radius $R$ this yields:
\begin{equation}
p_{nlj} \propto \int_0^R R_{nl_{j=1/2}}(r) \mathrm{d} r, 
\end{equation}
with $R_{nl_{j=1/2}}$ the radial wave-function of the large (small) part of the one electron orbital with quantum numbers $ns$ ($np_{1/2}$).
The relative matrix elements are:
\begin{equation}
\label{eq:pn}
\begin{array}{rlcrl}
p_{1s}&1 &\,\,\,&&\\
p_{2s}& 0.3669 &&p_{2p_{1/2}}& 0.0803\\
p_{3s}& 0.1712 &&p_{3p_{1/2}}& 0.0395\\
p_{4s}& 0.0842 &&p_{4p_{1/2}}& 0.0191\\
p_{5s}& 0.0338 &&p_{5p_{1/2}}& 0.0069\\
p_{6s}& 0.0095 &&&\\
\end{array}
\end{equation}
normalized to the capture probability of the 1s shell.

\section{\label{calc:numerics}Numerical stability and Block Lanczos}

Using a Lanczos algorithm we determine the $^{163}$Ho ground state before the electron capture event $\ket{\psi^{e^{-}}_{\text{Ho}}}$ using a multi-configurational representation. After finding the ground-state of the Ho atom we are able to calculate the deexcitation spectrum by looking at the time evolution of $T_{e^-} \ket{\psi^{e^{-}}_{\text{Ho}}}$ in the electronic potential of $^{163}$Dy using (\ref{eqSpectrum}). This accounts for the fact that the holes are created in the $^{163}$Ho ground state but the deexcitation energies are those of the Dy Hamiltonian with $Z=66$. Note that in a many body language it is actually not the Hamiltonian that changes during the electron capture event, but the Fock space the many body Hamiltonian acts upon changes. The peak positions and additional structures in the spectrum are directly encoded in the many-body Hamiltonian. The peak intensities are given by the transition operator $T_{e^-}$ and the interference between electron capture channels, multiplet formation and Auger decay. Both the intensity and peak energy are calculated without experimental input. As a consequence, the only parameters that cannot be calculated a priori within this approach are the Q-value, the total amplitude of the spectrum (half life of $^{163}$Ho) and, due to current restrictions of the basis set, the width of the peaks.

As $T$ acts on the $^{163}$Ho ground state, we obtain a linear combination of $2_s\times 10_i$ states which have one hole in each of the inner shells (the subscripts denote 2$_s$ spin states and 10$_i$ inner shells from which electron capture is possible). The energies of these holes vary widely between 16 eV for the 6s shell and 53 keV for 1s shell. This leads to numerical instabilities (number loss) if the resolvent in Eq. (\ref{eqSpectrum}) is calculated directly including all states. On computers with finite numerical accuracy it is necessary to separate the different energy scales. We achieve a seperation of energy scales and numerical stability with the use of a Block Lanczos routine. The starting vectors of our Block algorithm are the 10 states created by acting with each term in the operator $T_{e^-}$ on $\ket{\psi^{e^{-}}_{\text{Ho}}}$ separately. The resulting Green's function in this basis is represented by a 10 by 10 matrix. The full electron capture spectrum is given by the sum of all elements in this matrix, including the off diagonal terms.

In addition to numerical stability, the Block Lanczos routine has two other advantages. The first is that we do not need to sum explicitly over all possible one- or two-hole excited states, as their contributions to the spectrum appear naturally when the Krylov space is built up by the Lanczos algorithm. The second advantage is that we can easily restrict the Krylov space in order to study the contribution of certain states. In Figures \ref{figNTri1} and \ref{figMultiHoles} only the starting vectors and their matrix elements of the Hamiltonian have been used. Thus, the contribution of single holes sitting in the potential of the surrounding electrons is separated from the other effects like the hole scattering into a different orbital. To include the latter effect, we expand our Krylov space to $2_s\times 10_i\times 100$ states, where the 4f shell is restricted to have eleven electrons. In this setting we obtain the black spectrum in Fig. \ref{figRest11}, which modifies Fig. \ref{figNTri1} but still neglects Auger decays and the corresponding double vacancy excitations. These emerge if we remove the restriction on the 4f shell to have eleven electrons. All of these restrictions on the Krylov space are directly related to the restrictions on the Hilbert space discussed in Section \ref{res}.

\section{\label{sec:mixing} Hole mixing of principle quantum numbers}

\begin{figure*}[]
  \subfloat{\includegraphics[width=0.5\textwidth]{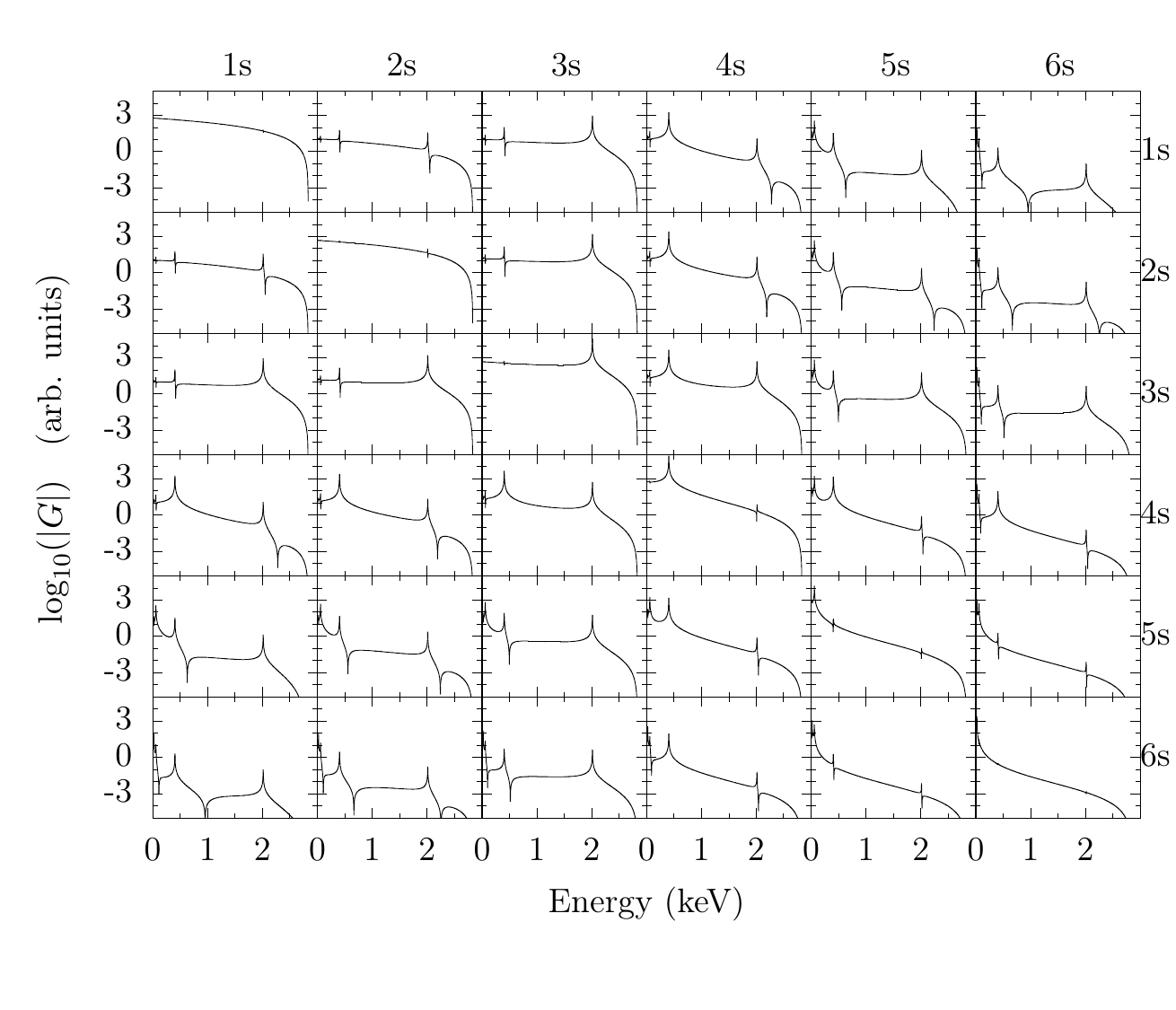}}
  \hfill
  \subfloat{\includegraphics[width=0.5\textwidth]{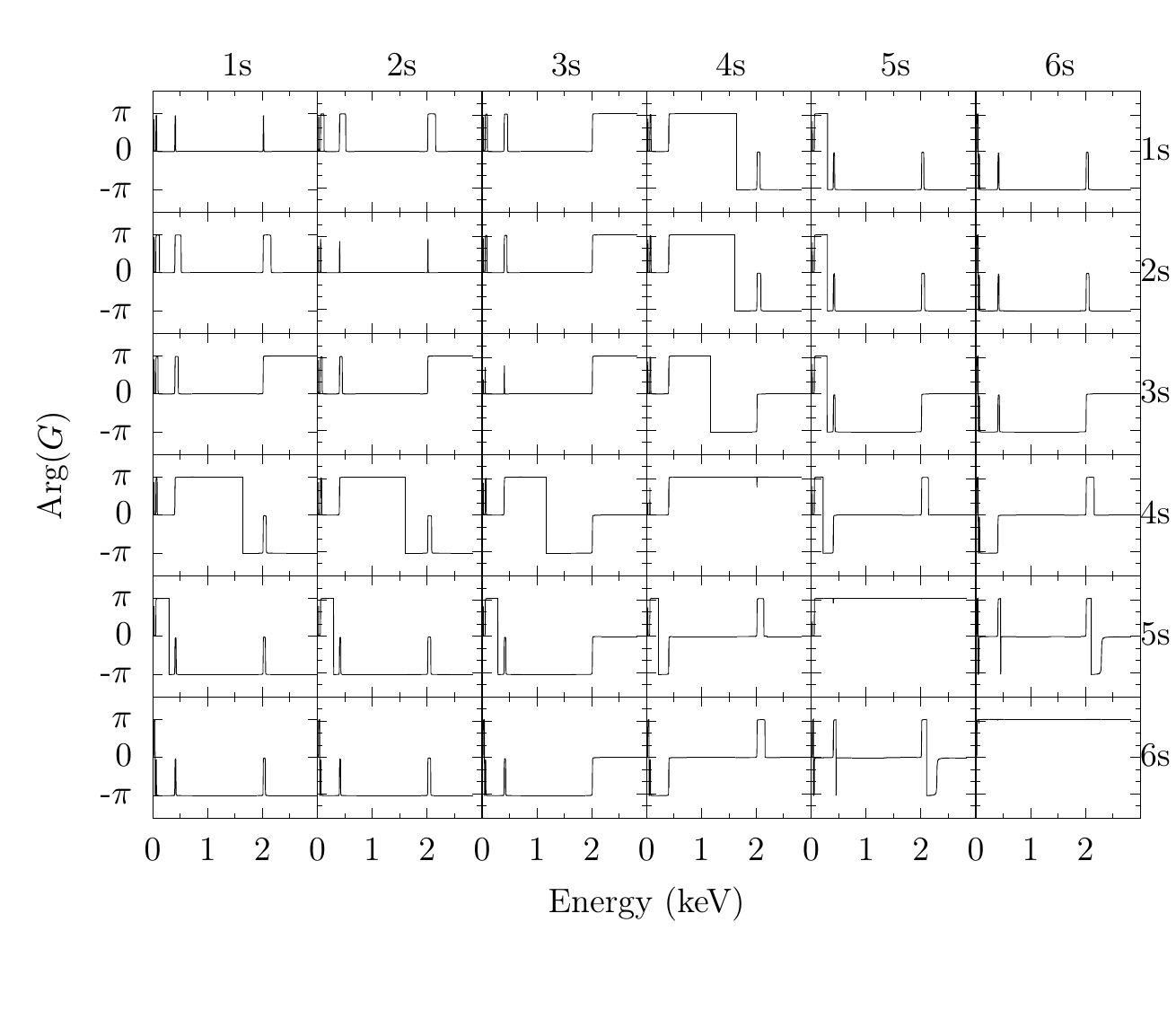}}
  \caption{\label{figMultiHoles}Norm and phase of the Green's function matrix $G_{ns,n's}=\braket{\psi_{\Ho}^{e^-}|c_{ns}^{\dag} \left((\omega-H_{\text{Dy}}+i\gamma /2)^{-1} - (\omega + H_{\text{Dy}}-i\gamma /2)^{-1}\right)  c_{n's}|\psi_{\Ho}^{e^-}}$. Diagonal elements show the spectrum resulting from the electron capture in the $ns=n's$ shell. Capture events from the $ns$ shell show resonances at all $n's$ binding energies. Off-diagonal elements show interference effects, i.e. how an electron annihilated in the $ns$ shell can be recreated in the $n's$ shell. For the calculation of the total intensity, the phase obtained during the scattering process is important, leading to Fano-like \cite{Nozieres:1969fk, Doniach:1970wr, Cornaglia:2007kn, Haverkort:2014hq} line-shapes.}
\end{figure*}

After a capture event from, for example, the $1s$ orbital, the nuclear potential changes. As the eigen-orbitals of $\psi_{\Ho}^{e^-}$ are different from the eigen-orbitals of $\psi_{\Dy}^{e^-}$, the $1s$ orbital in the potential of Ho has an overlap with all orbitals of $s$ character in the potential of Dy. This leads to so-called overlap and exchange corrections \cite{Bambynek:1977wz}. Where the overlap correction changes the intensity of each of the edges, the exchange correction leads to inferences between the edges.

For capture events in the $s$ shell, the impact of the hole-mixing is shown in Fig. \ref{figMultiHoles}. These spectra show the one particle Green's function (or propagator) after the creation of a core hole in the $ns$ shell. The diagonal panels show those functions where the hole is created and annihilated in the same shell, and the off-diagonal elements show the functions where the hole is recreated in a different shell from which it was annihilated. The overlap corrections change the peak height of the diagonal terms at the energy of the resonance where the core hole was created, and the exchange corrections lead to peaks at the other binding energies for the diagonal spectra.

It is important to realize that the so-called exchange interactions \cite{Bambynek:1977wz} come with a phase that changes across the resonance. The two major panels in Fig. \ref{figMultiHoles} show the norm and phase of $G(ns,n's)$. The question of whether there is constructive or destructive interference thus depends on the energy one considers and should not be treated as a constant scaling of the peak intensity. Even more important are the off-diagonal elements in $G(ns,n's)$. The measured intensity related to this Green's function matrix is proportional to $-\mathrm{Im}\sum_{n,n'} p_{ns} p_{n's} G(ns,n's,\omega)$, where $p_{ns}$ is the fractional capture probability amplitude as defined in Eq. \ref{eq:pn}. The off-diagonal interference terms enter with relatively large capture probabilities and are the main cause of the shift of intensity. As the interference terms enter with a phase, the resonances obtain Fano-like asymmetric line-shapes \cite{Nozieres:1969fk, Doniach:1970wr, Cornaglia:2007kn, Haverkort:2014hq}, which are important if one is interested in the tails of the spectrum.

To quantify the off-diagonal elements further, one can investigate the Hamiltonian $H_{\Dy}$ on a basis of the states $c_{nlj}\ket{\psi_{\Ho}^{e^-}}$, with $nlj\in \lbrace \text{1s - 6s, 2p$_{1/2}$ -5p$_{1/2}$} \rbrace$. 
\begin{widetext}
\begin{multline}
\label{eq:HDy}
\bf{H_{Dy}}=\\
\begin{blockarray}{ccccccccccccccccccccr}
\BAmulticolumn{2}{c}{c_{1s}\ket{\Psi_{\text{Ho}}}}& \BAmulticolumn{2}{c}{c_{2s}\ket{\Psi_{\text{Ho}}}}& \BAmulticolumn{2}{c}{c_{3s}\ket{\Psi_{\text{Ho}}}}& \BAmulticolumn{2}{c}{c_{4s}\ket{\Psi_{\text{Ho}}}}& \BAmulticolumn{2}{c}{c_{5s}\ket{\Psi_{\text{Ho}}}}& \BAmulticolumn{2}{c}{c_{6s}\ket{\Psi_{\text{Ho}}}}& \BAmulticolumn{2}{c}{c_{2p_{\frac{1}{2}}}\ket{\Psi_{\text{Ho}}}}& \BAmulticolumn{2}{c}{c_{3p_{\frac{1}{2}}}\ket{\Psi_{\text{Ho}}}}& \BAmulticolumn{2}{c}{c_{4p_{\frac{1}{2}}}\ket{\Psi_{\text{Ho}}}}& \BAmulticolumn{2}{c}{c_{5p_{\frac{1}{2}}}\ket{\Psi_{\text{Ho}}}}& \\
\begin{block}{(r@{.}lr@{.}lr@{.}lr@{.}lr@{.}lr@{.}lr@{.}lr@{.}lr@{.}lr@{.}l)r}
 53912&60 & -451&69 & -196&95 & -94&82 & -37&77 & -10&37 & 0&0& 0&0& 0&0& 0&0&\,\,c_{1s}\ket{\Psi_{\text{Ho}}} \\
 -451&69 & 9008&98 & -131&76 & -58&18 & -22&60 & -6&18 & 0&0& 0&0& 0&0& 0&0&\,\,c_{2s}\ket{\Psi_{\text{Ho}}} \\
 -196&95 & -131&76 & 2014&23 & -43&18 & -15&04 & -4&06 & 0&0& 0&0& 0&0& 0&0&\,\,c_{3s}\ket{\Psi_{\text{Ho}}} \\
 -94&82 & -58&18 & -43&18 & 403&24 & -15&38 & -3&97 & 0&0& 0&0& 0&0& 0&0&\,\,c_{4s}\ket{\Psi_{\text{Ho}}} \\
 -37&77 & -22&60 & -15&04 & -15&38 & 57&96 & -4&38 & 0&0& 0&0& 0&0& 0&0&\,\,c_{5s}\ket{\Psi_{\text{Ho}}} \\
 -10&37 & -6&18 & -4&06 & -3&97 & -4&38 & 21&63 & 0&0& 0&0& 0&0& 0&0&\,\,c_{6s}\ket{\Psi_{\text{Ho}}} \\
 0&0& 0&0& 0&0& 0&0& 0&0& 0&0& 8528&50 & -124&94 & -52&81 & -18&29 &\,\,c_{2p_{\frac{1}{2}}}\ket{\Psi_{\text{Ho}}} \\
 0&0& 0&0& 0&0& 0&0& 0&0& 0&0& -124&94 & 1808&28 & -44&04 & -13&59 &\,\,c_{3p_{\frac{1}{2}}}\ket{\Psi_{\text{Ho}}} \\
 0&0& 0&0& 0&0& 0&0& 0&0& 0&0& -52&81 & -44&04 & 325&31 & -14&18 &\,\,c_{4p_{\frac{1}{2}}}\ket{\Psi_{\text{Ho}}} \\
 0&0& 0&0& 0&0& 0&0& 0&0& 0&0& -18&29 & -13&59 & -14&18 & 37&03 &\,\,c_{5p_{\frac{1}{2}}}\ket{\Psi_{\text{Ho}}} \\
 \end{block}
\end{blockarray}
\end{multline}
\end{widetext}
This $10$ by $10$ matrix defines $G(nlj,n'lj,\omega)$ to first order in the Krylov basis expansion as
\begin{equation}
\mathbf{G}(\omega)=\frac{1}{\omega+\mathrm{i}\frac{\gamma}{2}-\mathbf{H_{Dy}}}-\frac{1}{\omega+\mathrm{i}\frac{\gamma}{2}+\mathbf{H_{Dy}}},
\end{equation}
where here $\mathbf{H_{Dy}}$ represents the ten by ten matrix created by evaluating the full Hamiltonian ($H_{\Dy}$) on a basis of the states where one core hole is created as given in Eq. \ref{eq:HDy}.

The off-diagonal elements are at maximum only a few percent of the energy difference between the states they couple, which explains the maximal energy shift of only a few eV and the relatively modest intensity transfer between the resonances when comparing the spectra shown in Figs. \ref{figNTri1} and \ref{figRest11}. For the understanding of the line-shape, the inference terms can become crucial once realistic core hole lifetimes are included due to decay into continuum states. 

\section{Relativistic corrections beyond the Dirac equation from quantum electrodynamics}
\begin{figure}[t]
\includegraphics[width=\columnwidth]{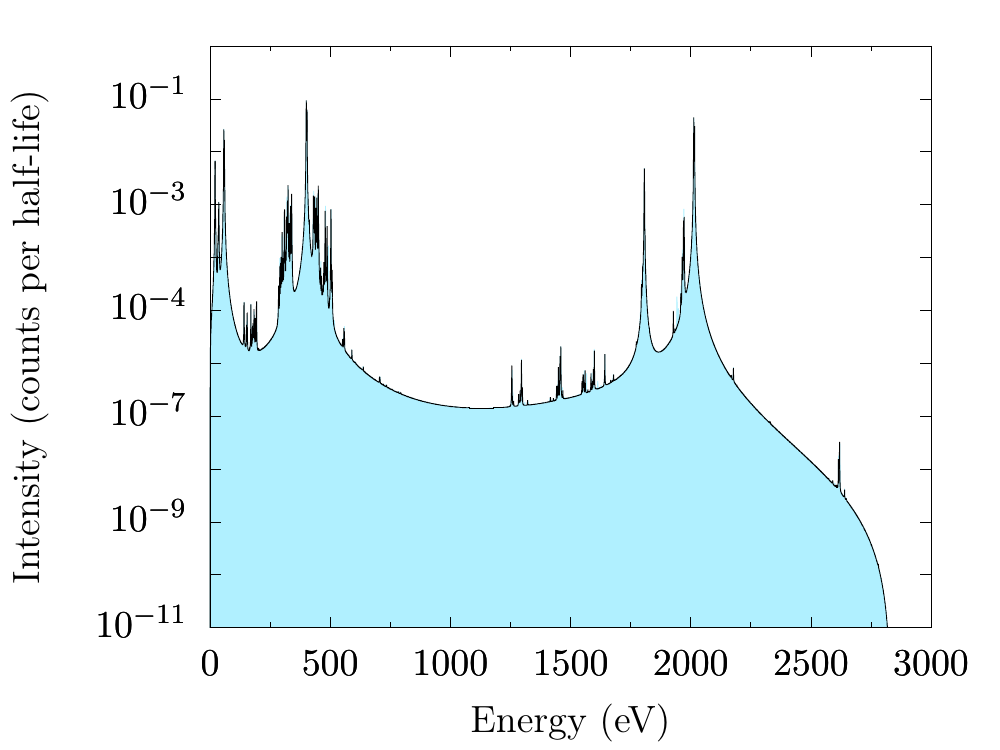}
\caption{\label{figBreitOff} Theoretical electron capture spectrum with (black) and without (blue) Breit interaction terms for shell occupation conserving scattering. Only a small amount of spectral weight is shifted between the peaks. }
\end{figure}
The discussed calculations include the Dirac-Coulomb Hamiltonian and first order corrections due to quantum electrodynamics. These corrections are the Breit interaction term for the Coulomb repulsion. The density functional theory calculations for the orbital wave functions defining our one particle basis set were done relativistically including the Breit interaction. The interaction term in our Hamiltonian also contains the Breit interaction but on the level of shell occupation conserving scattering only. To test importance of the level of corrections beyond the Dirac equation due to quantum electrodynamics we compare in Fig. \ref{figBreitOff} calculations including the Breit terms to calculations where the Breit term is neglected. Only very small changes are observed and further relativistic corrections due to quantum electrodynamics are assumed to be negligible.


\begin{thebibliography}{29}%
\makeatletter
\providecommand \@ifxundefined [1]{%
 \@ifx{#1\undefined}
}%
\providecommand \@ifnum [1]{%
 \ifnum #1\expandafter \@firstoftwo
 \else \expandafter \@secondoftwo
 \fi
}%
\providecommand \@ifx [1]{%
 \ifx #1\expandafter \@firstoftwo
 \else \expandafter \@secondoftwo
 \fi
}%
\providecommand \natexlab [1]{#1}%
\providecommand \enquote  [1]{``#1''}%
\providecommand \bibnamefont  [1]{#1}%
\providecommand \bibfnamefont [1]{#1}%
\providecommand \citenamefont [1]{#1}%
\providecommand \href@noop [0]{\@secondoftwo}%
\providecommand \href [0]{\begingroup \@sanitize@url \@href}%
\providecommand \@href[1]{\@@startlink{#1}\@@href}%
\providecommand \@@href[1]{\endgroup#1\@@endlink}%
\providecommand \@sanitize@url [0]{\catcode `\\12\catcode `\$12\catcode
  `\&12\catcode `\#12\catcode `\^12\catcode `\_12\catcode `\%12\relax}%
\providecommand \@@startlink[1]{}%
\providecommand \@@endlink[0]{}%
\providecommand \url  [0]{\begingroup\@sanitize@url \@url }%
\providecommand \@url [1]{\endgroup\@href {#1}{\urlprefix }}%
\providecommand \urlprefix  [0]{URL }%
\providecommand \Eprint [0]{\href }%
\providecommand \doibase [0]{http://dx.doi.org/}%
\providecommand \selectlanguage [0]{\@gobble}%
\providecommand \bibinfo  [0]{\@secondoftwo}%
\providecommand \bibfield  [0]{\@secondoftwo}%
\providecommand \translation [1]{[#1]}%
\providecommand \BibitemOpen [0]{}%
\providecommand \bibitemStop [0]{}%
\providecommand \bibitemNoStop [0]{.\EOS\space}%
\providecommand \EOS [0]{\spacefactor3000\relax}%
\providecommand \BibitemShut  [1]{\csname bibitem#1\endcsname}%
\let\auto@bib@innerbib\@empty
\bibitem [{\citenamefont {Drexlin}\ \emph {et~al.}(2013)\citenamefont
  {Drexlin}, \citenamefont {Hannen}, \citenamefont {Mertens},\ and\
  \citenamefont {Weinheimer}}]{Drexlin:2013jn}%
  \BibitemOpen
  \bibfield  {author} {\bibinfo {author} {\bibfnamefont {G.}~\bibnamefont
  {Drexlin}}, \bibinfo {author} {\bibfnamefont {V.}~\bibnamefont {Hannen}},
  \bibinfo {author} {\bibfnamefont {S.}~\bibnamefont {Mertens}}, \ and\
  \bibinfo {author} {\bibfnamefont {C.}~\bibnamefont {Weinheimer}},\
  }\href@noop {} {\bibfield  {journal} {\bibinfo  {journal} {Advances in High
  Energy Physics}\ }\textbf {\bibinfo {volume} {2013}},\ \bibinfo {pages} {1}
  (\bibinfo {year} {2013})}\BibitemShut {NoStop}%
\bibitem [{\citenamefont {Eliseev}\ \emph {et~al.}(2015)\citenamefont
  {Eliseev}, \citenamefont {Blaum}, \citenamefont {Block}, \citenamefont
  {Chenmarev}, \citenamefont {Dorrer}, \citenamefont {D{\"u}llmann},
  \citenamefont {Enss}, \citenamefont {Filianin}, \citenamefont {Gastaldo},
  \citenamefont {Goncharov}, \citenamefont {K{\"o}ster}, \citenamefont
  {Lautenschl{\"a}ger}, \citenamefont {Novikov}, \citenamefont {Rischka},
  \citenamefont {Sch{\"u}ssler}, \citenamefont {Schweikhard},\ and\
  \citenamefont {T{\"u}rler}}]{Eliseev:2015cp}%
  \BibitemOpen
  \bibfield  {author} {\bibinfo {author} {\bibfnamefont {S.}~\bibnamefont
  {Eliseev}}, \bibinfo {author} {\bibfnamefont {K.}~\bibnamefont {Blaum}},
  \bibinfo {author} {\bibfnamefont {M.}~\bibnamefont {Block}}, \bibinfo
  {author} {\bibfnamefont {S.}~\bibnamefont {Chenmarev}}, \bibinfo {author}
  {\bibfnamefont {H.}~\bibnamefont {Dorrer}}, \bibinfo {author} {\bibfnamefont
  {C.~E.}\ \bibnamefont {D{\"u}llmann}}, \bibinfo {author} {\bibfnamefont
  {C.}~\bibnamefont {Enss}}, \bibinfo {author} {\bibfnamefont {P.~E.}\
  \bibnamefont {Filianin}}, \bibinfo {author} {\bibfnamefont {L.}~\bibnamefont
  {Gastaldo}}, \bibinfo {author} {\bibfnamefont {M.}~\bibnamefont {Goncharov}},
  \bibinfo {author} {\bibfnamefont {U.}~\bibnamefont {K{\"o}ster}}, \bibinfo
  {author} {\bibfnamefont {F.}~\bibnamefont {Lautenschl{\"a}ger}}, \bibinfo
  {author} {\bibfnamefont {Y.~N.}\ \bibnamefont {Novikov}}, \bibinfo {author}
  {\bibfnamefont {A.}~\bibnamefont {Rischka}}, \bibinfo {author} {\bibfnamefont
  {R.~X.}\ \bibnamefont {Sch{\"u}ssler}}, \bibinfo {author} {\bibfnamefont
  {L.}~\bibnamefont {Schweikhard}}, \ and\ \bibinfo {author} {\bibfnamefont
  {A.}~\bibnamefont {T{\"u}rler}},\ }\href@noop {} {\bibfield  {journal}
  {\bibinfo  {journal} {Phys. Rev. Lett.}\ }\textbf {\bibinfo {volume} {115}},\
  \bibinfo {pages} {062501} (\bibinfo {year} {2015})}\BibitemShut {NoStop}%
\bibitem [{\citenamefont {De~R{\'u}jula}\ and\ \citenamefont
  {Lusignoli}(1982)}]{DeRujula:1982us}%
  \BibitemOpen
  \bibfield  {author} {\bibinfo {author} {\bibfnamefont {A.}~\bibnamefont
  {De~R{\'u}jula}}\ and\ \bibinfo {author} {\bibfnamefont {M.}~\bibnamefont
  {Lusignoli}},\ }\href@noop {} {\bibfield  {journal} {\bibinfo  {journal}
  {Physics Letters B}\ }\textbf {\bibinfo {volume} {118}},\ \bibinfo {pages}
  {429} (\bibinfo {year} {1982})}\BibitemShut {NoStop}%
\bibitem [{\citenamefont {Gastaldo}\ \emph {et~al.}(2017)\citenamefont
  {Gastaldo}, \citenamefont {Blaum}, \citenamefont {Chrysalidis}, \citenamefont
  {Day~Goodacre}, \citenamefont {Domula}, \citenamefont {Door}, \citenamefont
  {Dorrer}, \citenamefont {D~llmann}, \citenamefont {Eberhardt}, \citenamefont
  {Eliseev}, \citenamefont {Enss}, \citenamefont {Faessler}, \citenamefont
  {Filianin}, \citenamefont {Fleischmann}, \citenamefont {Fonnesu},
  \citenamefont {Gamer}, \citenamefont {Haas}, \citenamefont {Hassel},
  \citenamefont {Hengstler}, \citenamefont {Jochum}, \citenamefont {Johnston},
  \citenamefont {Kebschull}, \citenamefont {Kempf}, \citenamefont {Kieck},
  \citenamefont {K~ster}, \citenamefont {Lahiri}, \citenamefont {Maiti},
  \citenamefont {Mantegazzini}, \citenamefont {Marsh}, \citenamefont
  {Neroutsos}, \citenamefont {Novikov}, \citenamefont {Ranitzsch},
  \citenamefont {Rothe}, \citenamefont {Rischka}, \citenamefont {Saenz},
  \citenamefont {Sander}, \citenamefont {Schneider}, \citenamefont {Scholl},
  \citenamefont {Sch~ssler}, \citenamefont {Schweiger}, \citenamefont {{\v
  S}imkovic}, \citenamefont {Stora}, \citenamefont {Sz~cs}, \citenamefont
  {T~rler}, \citenamefont {Veinhard}, \citenamefont {Weber}, \citenamefont
  {Wegner}, \citenamefont {Wendt},\ and\ \citenamefont
  {Zuber}}]{Gastaldo:2017ch}%
  \BibitemOpen
  \bibfield  {author} {\bibinfo {author} {\bibfnamefont {L.}~\bibnamefont
  {Gastaldo}}, \bibinfo {author} {\bibfnamefont {K.}~\bibnamefont {Blaum}},
  \bibinfo {author} {\bibfnamefont {K.}~\bibnamefont {Chrysalidis}}, \bibinfo
  {author} {\bibfnamefont {T.}~\bibnamefont {Day~Goodacre}}, \bibinfo {author}
  {\bibfnamefont {A.}~\bibnamefont {Domula}}, \bibinfo {author} {\bibfnamefont
  {M.}~\bibnamefont {Door}}, \bibinfo {author} {\bibfnamefont {H.}~\bibnamefont
  {Dorrer}}, \bibinfo {author} {\bibfnamefont {C.~E.}\ \bibnamefont
  {D~llmann}}, \bibinfo {author} {\bibfnamefont {K.}~\bibnamefont {Eberhardt}},
  \bibinfo {author} {\bibfnamefont {S.}~\bibnamefont {Eliseev}}, \bibinfo
  {author} {\bibfnamefont {C.}~\bibnamefont {Enss}}, \bibinfo {author}
  {\bibfnamefont {A.}~\bibnamefont {Faessler}}, \bibinfo {author}
  {\bibfnamefont {P.}~\bibnamefont {Filianin}}, \bibinfo {author}
  {\bibfnamefont {A.}~\bibnamefont {Fleischmann}}, \bibinfo {author}
  {\bibfnamefont {D.}~\bibnamefont {Fonnesu}}, \bibinfo {author} {\bibfnamefont
  {L.}~\bibnamefont {Gamer}}, \bibinfo {author} {\bibfnamefont
  {R.}~\bibnamefont {Haas}}, \bibinfo {author} {\bibfnamefont {C.}~\bibnamefont
  {Hassel}}, \bibinfo {author} {\bibfnamefont {D.}~\bibnamefont {Hengstler}},
  \bibinfo {author} {\bibfnamefont {J.}~\bibnamefont {Jochum}}, \bibinfo
  {author} {\bibfnamefont {K.}~\bibnamefont {Johnston}}, \bibinfo {author}
  {\bibfnamefont {U.}~\bibnamefont {Kebschull}}, \bibinfo {author}
  {\bibfnamefont {S.}~\bibnamefont {Kempf}}, \bibinfo {author} {\bibfnamefont
  {T.}~\bibnamefont {Kieck}}, \bibinfo {author} {\bibfnamefont
  {U.}~\bibnamefont {K~ster}}, \bibinfo {author} {\bibfnamefont
  {S.}~\bibnamefont {Lahiri}}, \bibinfo {author} {\bibfnamefont
  {M.}~\bibnamefont {Maiti}}, \bibinfo {author} {\bibfnamefont
  {F.}~\bibnamefont {Mantegazzini}}, \bibinfo {author} {\bibfnamefont
  {B.}~\bibnamefont {Marsh}}, \bibinfo {author} {\bibfnamefont
  {P.}~\bibnamefont {Neroutsos}}, \bibinfo {author} {\bibfnamefont {Y.~N.}\
  \bibnamefont {Novikov}}, \bibinfo {author} {\bibfnamefont {P.~C.~O.}\
  \bibnamefont {Ranitzsch}}, \bibinfo {author} {\bibfnamefont {S.}~\bibnamefont
  {Rothe}}, \bibinfo {author} {\bibfnamefont {A.}~\bibnamefont {Rischka}},
  \bibinfo {author} {\bibfnamefont {A.}~\bibnamefont {Saenz}}, \bibinfo
  {author} {\bibfnamefont {O.}~\bibnamefont {Sander}}, \bibinfo {author}
  {\bibfnamefont {F.}~\bibnamefont {Schneider}}, \bibinfo {author}
  {\bibfnamefont {S.}~\bibnamefont {Scholl}}, \bibinfo {author} {\bibfnamefont
  {R.~X.}\ \bibnamefont {Sch~ssler}}, \bibinfo {author} {\bibfnamefont
  {C.}~\bibnamefont {Schweiger}}, \bibinfo {author} {\bibfnamefont
  {F.}~\bibnamefont {{\v S}imkovic}}, \bibinfo {author} {\bibfnamefont
  {T.}~\bibnamefont {Stora}}, \bibinfo {author} {\bibfnamefont
  {Z.}~\bibnamefont {Sz~cs}}, \bibinfo {author} {\bibfnamefont
  {A.}~\bibnamefont {T~rler}}, \bibinfo {author} {\bibfnamefont
  {M.}~\bibnamefont {Veinhard}}, \bibinfo {author} {\bibfnamefont
  {M.}~\bibnamefont {Weber}}, \bibinfo {author} {\bibfnamefont
  {M.}~\bibnamefont {Wegner}}, \bibinfo {author} {\bibfnamefont
  {K.}~\bibnamefont {Wendt}}, \ and\ \bibinfo {author} {\bibfnamefont
  {K.}~\bibnamefont {Zuber}},\ }\href@noop {} {\bibfield  {journal} {\bibinfo
  {journal} {Eur. Phys. J. Spec. Top.}\ }\textbf {\bibinfo {volume} {226}},\
  \bibinfo {pages} {1623} (\bibinfo {year} {2017})}\BibitemShut {NoStop}%
\bibitem [{\citenamefont {Alpert}\ \emph {et~al.}(2015)\citenamefont {Alpert},
  \citenamefont {Balata}, \citenamefont {Bennett}, \citenamefont {Biasotti},
  \citenamefont {Boragno}, \citenamefont {Brofferio}, \citenamefont {Ceriale},
  \citenamefont {Corsini}, \citenamefont {Day}, \citenamefont {De~Gerone},
  \citenamefont {Dressler}, \citenamefont {Faverzani}, \citenamefont {Ferri},
  \citenamefont {Fowler}, \citenamefont {Gatti}, \citenamefont {Giachero},
  \citenamefont {Hays-Wehle}, \citenamefont {Heinitz}, \citenamefont {Hilton},
  \citenamefont {K{\"o}ster}, \citenamefont {Lusignoli}, \citenamefont {Maino},
  \citenamefont {Mates}, \citenamefont {Nisi}, \citenamefont {Nizzolo},
  \citenamefont {Nucciotti}, \citenamefont {Pessina}, \citenamefont
  {Pizzigoni}, \citenamefont {Puiu}, \citenamefont {Ragazzi}, \citenamefont
  {Reintsema}, \citenamefont {Gomes}, \citenamefont {Schmidt}, \citenamefont
  {Schumann}, \citenamefont {Sisti}, \citenamefont {Swetz}, \citenamefont
  {Terranova},\ and\ \citenamefont {Ullom}}]{Alpert:2015gi}%
  \BibitemOpen
  \bibfield  {author} {\bibinfo {author} {\bibfnamefont {B.}~\bibnamefont
  {Alpert}}, \bibinfo {author} {\bibfnamefont {M.}~\bibnamefont {Balata}},
  \bibinfo {author} {\bibfnamefont {D.}~\bibnamefont {Bennett}}, \bibinfo
  {author} {\bibfnamefont {M.}~\bibnamefont {Biasotti}}, \bibinfo {author}
  {\bibfnamefont {C.}~\bibnamefont {Boragno}}, \bibinfo {author} {\bibfnamefont
  {C.}~\bibnamefont {Brofferio}}, \bibinfo {author} {\bibfnamefont
  {V.}~\bibnamefont {Ceriale}}, \bibinfo {author} {\bibfnamefont
  {D.}~\bibnamefont {Corsini}}, \bibinfo {author} {\bibfnamefont {P.~K.}\
  \bibnamefont {Day}}, \bibinfo {author} {\bibfnamefont {M.}~\bibnamefont
  {De~Gerone}}, \bibinfo {author} {\bibfnamefont {R.}~\bibnamefont {Dressler}},
  \bibinfo {author} {\bibfnamefont {M.}~\bibnamefont {Faverzani}}, \bibinfo
  {author} {\bibfnamefont {E.}~\bibnamefont {Ferri}}, \bibinfo {author}
  {\bibfnamefont {J.}~\bibnamefont {Fowler}}, \bibinfo {author} {\bibfnamefont
  {F.}~\bibnamefont {Gatti}}, \bibinfo {author} {\bibfnamefont
  {A.}~\bibnamefont {Giachero}}, \bibinfo {author} {\bibfnamefont
  {J.}~\bibnamefont {Hays-Wehle}}, \bibinfo {author} {\bibfnamefont
  {S.}~\bibnamefont {Heinitz}}, \bibinfo {author} {\bibfnamefont
  {G.}~\bibnamefont {Hilton}}, \bibinfo {author} {\bibfnamefont
  {U.}~\bibnamefont {K{\"o}ster}}, \bibinfo {author} {\bibfnamefont
  {M.}~\bibnamefont {Lusignoli}}, \bibinfo {author} {\bibfnamefont
  {M.}~\bibnamefont {Maino}}, \bibinfo {author} {\bibfnamefont
  {J.}~\bibnamefont {Mates}}, \bibinfo {author} {\bibfnamefont
  {S.}~\bibnamefont {Nisi}}, \bibinfo {author} {\bibfnamefont {R.}~\bibnamefont
  {Nizzolo}}, \bibinfo {author} {\bibfnamefont {A.}~\bibnamefont {Nucciotti}},
  \bibinfo {author} {\bibfnamefont {G.}~\bibnamefont {Pessina}}, \bibinfo
  {author} {\bibfnamefont {G.}~\bibnamefont {Pizzigoni}}, \bibinfo {author}
  {\bibfnamefont {A.}~\bibnamefont {Puiu}}, \bibinfo {author} {\bibfnamefont
  {S.}~\bibnamefont {Ragazzi}}, \bibinfo {author} {\bibfnamefont
  {C.}~\bibnamefont {Reintsema}}, \bibinfo {author} {\bibfnamefont {M.~R.}\
  \bibnamefont {Gomes}}, \bibinfo {author} {\bibfnamefont {D.}~\bibnamefont
  {Schmidt}}, \bibinfo {author} {\bibfnamefont {D.}~\bibnamefont {Schumann}},
  \bibinfo {author} {\bibfnamefont {M.}~\bibnamefont {Sisti}}, \bibinfo
  {author} {\bibfnamefont {D.}~\bibnamefont {Swetz}}, \bibinfo {author}
  {\bibfnamefont {F.}~\bibnamefont {Terranova}}, \ and\ \bibinfo {author}
  {\bibfnamefont {J.}~\bibnamefont {Ullom}},\ }\href@noop {} {\bibfield
  {journal} {\bibinfo  {journal} {Eur. Phys. J. C}\ }\textbf {\bibinfo {volume}
  {75}},\ \bibinfo {pages} {27} (\bibinfo {year} {2015})}\BibitemShut {NoStop}%
\bibitem [{\citenamefont {Croce}\ \emph {et~al.}(2016)\citenamefont {Croce},
  \citenamefont {Hoover}, \citenamefont {Rabin}, \citenamefont {Bond},
  \citenamefont {Wolfsberg}, \citenamefont {Schmidt},\ and\ \citenamefont
  {Ullom}}]{Croce:2016dp}%
  \BibitemOpen
  \bibfield  {author} {\bibinfo {author} {\bibfnamefont {M.~P.}\ \bibnamefont
  {Croce}}, \bibinfo {author} {\bibfnamefont {A.~S.}\ \bibnamefont {Hoover}},
  \bibinfo {author} {\bibfnamefont {M.~W.}\ \bibnamefont {Rabin}}, \bibinfo
  {author} {\bibfnamefont {E.~M.}\ \bibnamefont {Bond}}, \bibinfo {author}
  {\bibfnamefont {L.~E.}\ \bibnamefont {Wolfsberg}}, \bibinfo {author}
  {\bibfnamefont {D.~R.}\ \bibnamefont {Schmidt}}, \ and\ \bibinfo {author}
  {\bibfnamefont {J.~N.}\ \bibnamefont {Ullom}},\ }\href@noop {} {\bibfield
  {journal} {\bibinfo  {journal} {Journal of Low Temperature Physics}\ }\textbf
  {\bibinfo {volume} {184}},\ \bibinfo {pages} {938} (\bibinfo {year}
  {2016})}\BibitemShut {NoStop}%
\bibitem [{\citenamefont {Enss}(2005)}]{Enss:2005ue}%
  \BibitemOpen
  \bibinfo {editor} {\bibfnamefont {C.}~\bibnamefont {Enss}},\ ed.,\ \href@noop
  {} {\emph {\bibinfo {title} {{Cryogenic Particle Detection}}}},\ \bibinfo
  {series} {Topics in Applied Physics}, Vol.~\bibinfo {volume} {99}\ (\bibinfo
  {publisher} {Springer},\ \bibinfo {year} {2005})\BibitemShut {NoStop}%
\bibitem [{\citenamefont {Faessler}\ \emph {et~al.}(2015)\citenamefont
  {Faessler}, \citenamefont {Enss}, \citenamefont {Gastaldo},\ and\
  \citenamefont {{\v S}imkovic}}]{Faessler:2015dg}%
  \BibitemOpen
  \bibfield  {author} {\bibinfo {author} {\bibfnamefont {A.}~\bibnamefont
  {Faessler}}, \bibinfo {author} {\bibfnamefont {C.}~\bibnamefont {Enss}},
  \bibinfo {author} {\bibfnamefont {L.}~\bibnamefont {Gastaldo}}, \ and\
  \bibinfo {author} {\bibfnamefont {F.}~\bibnamefont {{\v S}imkovic}},\
  }\href@noop {} {\bibfield  {journal} {\bibinfo  {journal} {Phys. Rev. C}\
  }\textbf {\bibinfo {volume} {91}},\ \bibinfo {pages} {064302} (\bibinfo
  {year} {2015})}\BibitemShut {NoStop}%
\bibitem [{\citenamefont {Faessler}\ and\ \citenamefont {{\v
  S}imkovic}(2015)}]{Faessler:2015ck}%
  \BibitemOpen
  \bibfield  {author} {\bibinfo {author} {\bibfnamefont {A.}~\bibnamefont
  {Faessler}}\ and\ \bibinfo {author} {\bibfnamefont {F.}~\bibnamefont {{\v
  S}imkovic}},\ }\href@noop {} {\bibfield  {journal} {\bibinfo  {journal}
  {Phys. Rev. C}\ }\textbf {\bibinfo {volume} {91}},\ \bibinfo
  {pages} {045505}  (\bibinfo {year} {2015})}\BibitemShut {NoStop}%
\bibitem [{\citenamefont {Robertson}(2015)}]{Robertson:2015dg}%
  \BibitemOpen
  \bibfield  {author} {\bibinfo {author} {\bibfnamefont {R.~G.~H.}\
  \bibnamefont {Robertson}},\ }\href@noop {} {\bibfield  {journal} {\bibinfo
  {journal} {Phys. Rev. C}\ }\textbf {\bibinfo {volume} {91}},\ \bibinfo
  {pages} {035504} (\bibinfo {year} {2015})}\BibitemShut {NoStop}%
\bibitem [{\citenamefont {De~R{\'u}jula}\ and\ \citenamefont
  {Lusignoli}(2016)}]{DeRujula:2016cp}%
  \BibitemOpen
  \bibfield  {author} {\bibinfo {author} {\bibfnamefont {A.}~\bibnamefont
  {De~R{\'u}jula}}\ and\ \bibinfo {author} {\bibfnamefont {M.}~\bibnamefont
  {Lusignoli}},\ }\href@noop {} {\bibfield  {journal} {\bibinfo  {journal} {J.
  High Energ. Phys.}\ }\textbf {\bibinfo {volume} {2016}},\ \bibinfo {pages}
  {15} (\bibinfo {year} {2016})}\BibitemShut {NoStop}%
\bibitem [{\citenamefont {Faessler}\ \emph {et~al.}(2017)\citenamefont
  {Faessler}, \citenamefont {Gastaldo},\ and\ \citenamefont {{\v
  S}imkovic}}]{Faessler:2017hq}%
  \BibitemOpen
  \bibfield  {author} {\bibinfo {author} {\bibfnamefont {A.}~\bibnamefont
  {Faessler}}, \bibinfo {author} {\bibfnamefont {L.}~\bibnamefont {Gastaldo}},
  \ and\ \bibinfo {author} {\bibfnamefont {F.}~\bibnamefont {{\v S}imkovic}},\
  }\href@noop {} {\bibfield  {journal} {\bibinfo  {journal} {Phys. Rev. C}\
  }\textbf {\bibinfo {volume} {95}},\ \bibinfo {pages} {045502} (\bibinfo
  {year} {2017})}\BibitemShut {NoStop}%
\bibitem [{\citenamefont {de~Groot}\ and\ \citenamefont
  {Kotani}(2008)}]{deGroot:2008wo}%
  \BibitemOpen
  \bibfield  {author} {\bibinfo {author} {\bibfnamefont {F.~M.~F.}\
  \bibnamefont {de~Groot}}\ and\ \bibinfo {author} {\bibfnamefont
  {A.}~\bibnamefont {Kotani}},\ }\href@noop {} {\emph {\bibinfo {title} {{Core
  Level Spectroscopy of Solids}}}}\ (\bibinfo  {publisher} {CRC Press},\
  \bibinfo {year} {2008})\BibitemShut {NoStop}%
\bibitem [{\citenamefont {Tanaka}\ and\ \citenamefont
  {Jo}(1995)}]{Tanaka:1995tl}%
  \BibitemOpen
  \bibfield  {author} {\bibinfo {author} {\bibfnamefont {A.}~\bibnamefont
  {Tanaka}}\ and\ \bibinfo {author} {\bibfnamefont {T.}~\bibnamefont {Jo}},\
  }\href@noop {} {\bibfield  {journal} {\bibinfo  {journal} {J. Phys. Soc.
  Jpn.}\ } (\bibinfo {year} {1995})}\BibitemShut {NoStop}%
\bibitem [{\citenamefont {Rehr}\ \emph {et~al.}(2009)\citenamefont {Rehr},
  \citenamefont {Kas}, \citenamefont {Prange}, \citenamefont {Sorini},
  \citenamefont {Takimoto},\ and\ \citenamefont {Vila}}]{Rehr:2009eu}%
  \BibitemOpen
  \bibfield  {author} {\bibinfo {author} {\bibfnamefont {J.~J.}\ \bibnamefont
  {Rehr}}, \bibinfo {author} {\bibfnamefont {J.~J.}\ \bibnamefont {Kas}},
  \bibinfo {author} {\bibfnamefont {M.~P.}\ \bibnamefont {Prange}}, \bibinfo
  {author} {\bibfnamefont {A.~P.}\ \bibnamefont {Sorini}}, \bibinfo {author}
  {\bibfnamefont {Y.}~\bibnamefont {Takimoto}}, \ and\ \bibinfo {author}
  {\bibfnamefont {F.}~\bibnamefont {Vila}},\ }\href@noop {} {\bibfield
  {journal} {\bibinfo  {journal} {C. R. Physique}\ }\textbf {\bibinfo {volume}
  {10}},\ \bibinfo {pages} {548} (\bibinfo {year} {2009})}\BibitemShut
  {NoStop}%
\bibitem [{\citenamefont {Haverkort}\ \emph {et~al.}(2012)\citenamefont
  {Haverkort}, \citenamefont {Zwierzycki},\ and\ \citenamefont
  {Andersen}}]{Haverkort:2012du}%
  \BibitemOpen
  \bibfield  {author} {\bibinfo {author} {\bibfnamefont {M.~W.}\ \bibnamefont
  {Haverkort}}, \bibinfo {author} {\bibfnamefont {M.}~\bibnamefont
  {Zwierzycki}}, \ and\ \bibinfo {author} {\bibfnamefont {O.~K.}\ \bibnamefont
  {Andersen}},\ }\href@noop {} {\bibfield  {journal} {\bibinfo  {journal}
  {Phys. Rev. B}\ }\textbf {\bibinfo {volume} {85}},\ \bibinfo {pages} {165113}
  (\bibinfo {year} {2012})}\BibitemShut {NoStop}%
\bibitem [{\citenamefont {Haverkort}\ \emph {et~al.}(2014)\citenamefont
  {Haverkort}, \citenamefont {Sangiovanni}, \citenamefont {Hansmann},
  \citenamefont {Toschi}, \citenamefont {Lu},\ and\ \citenamefont
  {Macke}}]{Haverkort:2014hq}%
  \BibitemOpen
  \bibfield  {author} {\bibinfo {author} {\bibfnamefont {M.~W.}\ \bibnamefont
  {Haverkort}}, \bibinfo {author} {\bibfnamefont {G.}~\bibnamefont
  {Sangiovanni}}, \bibinfo {author} {\bibfnamefont {P.}~\bibnamefont
  {Hansmann}}, \bibinfo {author} {\bibfnamefont {A.}~\bibnamefont {Toschi}},
  \bibinfo {author} {\bibfnamefont {Y.}~\bibnamefont {Lu}}, \ and\ \bibinfo
  {author} {\bibfnamefont {S.}~\bibnamefont {Macke}},\ }\href@noop {}
  {\bibfield  {journal} {\bibinfo  {journal} {Europhys. Lett.}\ }\textbf
  {\bibinfo {volume} {108}},\ \bibinfo {pages} {57004} (\bibinfo {year}
  {2014})}\BibitemShut {NoStop}%
\bibitem [{\citenamefont {Bergmann}\ \emph {et~al.}(1999)\citenamefont
  {Bergmann}, \citenamefont {Glatzel}, \citenamefont {de~Groot},\ and\
  \citenamefont {Cramer}}]{Bergmann:1999dx}%
  \BibitemOpen
  \bibfield  {author} {\bibinfo {author} {\bibfnamefont {U.}~\bibnamefont
  {Bergmann}}, \bibinfo {author} {\bibfnamefont {P.}~\bibnamefont {Glatzel}},
  \bibinfo {author} {\bibfnamefont {F.~M.~F.}\ \bibnamefont {de~Groot}}, \ and\
  \bibinfo {author} {\bibfnamefont {S.~P.}\ \bibnamefont {Cramer}},\
  }\href@noop {} {\bibfield  {journal} {\bibinfo  {journal} {J. Am. Chem.
  Soc.}\ }\textbf {\bibinfo {volume} {121}},\ \bibinfo {pages} {4926} (\bibinfo
  {year} {1999})}\BibitemShut {NoStop}%
\bibitem [{\citenamefont {Glatzel}\ \emph {et~al.}(2001)\citenamefont
  {Glatzel}, \citenamefont {Bergmann}, \citenamefont {de~Groot},\ and\
  \citenamefont {Cramer}}]{Glatzel:2001ia}%
  \BibitemOpen
  \bibfield  {author} {\bibinfo {author} {\bibfnamefont {P.}~\bibnamefont
  {Glatzel}}, \bibinfo {author} {\bibfnamefont {U.}~\bibnamefont {Bergmann}},
  \bibinfo {author} {\bibfnamefont {F.~M.~F.}\ \bibnamefont {de~Groot}}, \ and\
  \bibinfo {author} {\bibfnamefont {S.~P.}\ \bibnamefont {Cramer}},\
  }\href@noop {} {\bibfield  {journal} {\bibinfo  {journal} {Phys. Rev. B}\
  }\textbf {\bibinfo {volume} {64}},\ \bibinfo {pages} {045109} (\bibinfo
  {year} {2001})}\BibitemShut {NoStop}%
\bibitem [{\citenamefont {Anderson}(1967)}]{Anderson:1967tr}%
  \BibitemOpen
  \bibfield  {author} {\bibinfo {author} {\bibfnamefont {P.~W.}\ \bibnamefont
  {Anderson}},\ }\href@noop {} {\bibfield  {journal} {\bibinfo  {journal}
  {Phys. Rev. Lett.}\ }\textbf {\bibinfo {volume} {18}},\ \bibinfo {pages}
  {1049} (\bibinfo {year} {1967})}\BibitemShut {NoStop}%
\bibitem [{\citenamefont {Ranitzsch}\ \emph {et~al.}(2017)\citenamefont
  {Ranitzsch}, \citenamefont {Hassel}, \citenamefont {Wegner}, \citenamefont
  {Hengstler}, \citenamefont {Kempf}, \citenamefont {Fleischmann},
  \citenamefont {Enss}, \citenamefont {Gastaldo}, \citenamefont {Herlert},\
  and\ \citenamefont {Johnston}}]{Ranitzsch:2017cp}%
  \BibitemOpen
  \bibfield  {author} {\bibinfo {author} {\bibfnamefont {P.~C.~-O.}\
  \bibnamefont {Ranitzsch}}, \bibinfo {author} {\bibfnamefont {C.}~\bibnamefont
  {Hassel}}, \bibinfo {author} {\bibfnamefont {M.}~\bibnamefont {Wegner}},
  \bibinfo {author} {\bibfnamefont {D.}~\bibnamefont {Hengstler}}, \bibinfo
  {author} {\bibfnamefont {S.}~\bibnamefont {Kempf}}, \bibinfo {author}
  {\bibfnamefont {A.}~\bibnamefont {Fleischmann}}, \bibinfo {author}
  {\bibfnamefont {C.}~\bibnamefont {Enss}}, \bibinfo {author} {\bibfnamefont
  {L.}~\bibnamefont {Gastaldo}}, \bibinfo {author} {\bibfnamefont
  {A.}~\bibnamefont {Herlert}}, \ and\ \bibinfo {author} {\bibfnamefont
  {K.}~\bibnamefont {Johnston}},\ }\href@noop {} {\bibfield  {journal}
  {\bibinfo  {journal} {Phys. Rev. Lett.}\ }\textbf {\bibinfo {volume} {119}},\
  \bibinfo {pages} {122501} (\bibinfo {year} {2017})}\BibitemShut {NoStop}%
\bibitem [{\citenamefont {Koepernik}\ and\ \citenamefont
  {Eschrig}(1999)}]{Koepernik:1999uw}%
  \BibitemOpen
  \bibfield  {author} {\bibinfo {author} {\bibfnamefont {K.}~\bibnamefont
  {Koepernik}}\ and\ \bibinfo {author} {\bibfnamefont {H.}~\bibnamefont
  {Eschrig}},\ }\href@noop {} {\bibfield  {journal} {\bibinfo  {journal} {Phys.
  Rev. B}\ }\textbf {\bibinfo {volume} {59}},\ \bibinfo {pages} {1743}
  (\bibinfo {year} {1999})}\BibitemShut {NoStop}%
\bibitem [{\citenamefont {Opahle}\ \emph {et~al.}(1999)\citenamefont {Opahle},
  \citenamefont {Koepernik},\ and\ \citenamefont {Eschrig}}]{Opahle:1999tx}%
  \BibitemOpen
  \bibfield  {author} {\bibinfo {author} {\bibfnamefont {I.}~\bibnamefont
  {Opahle}}, \bibinfo {author} {\bibfnamefont {K.}~\bibnamefont {Koepernik}}, \
  and\ \bibinfo {author} {\bibfnamefont {H.}~\bibnamefont {Eschrig}},\
  }\href@noop {} {\bibfield  {journal} {\bibinfo  {journal} {Phys. Rev. B}\
  }\textbf {\bibinfo {volume} {60}},\ \bibinfo {pages} {14035} (\bibinfo {year}
  {1999})}\BibitemShut {NoStop}%
\bibitem [{\citenamefont {Eschrig}\ \emph {et~al.}(2004)\citenamefont
  {Eschrig}, \citenamefont {Richter},\ and\ \citenamefont
  {Opahle}}]{Eschrig:2004wn}%
  \BibitemOpen
  \bibfield  {author} {\bibinfo {author} {\bibfnamefont {H.}~\bibnamefont
  {Eschrig}}, \bibinfo {author} {\bibfnamefont {M.}~\bibnamefont {Richter}}, \
  and\ \bibinfo {author} {\bibfnamefont {I.}~\bibnamefont {Opahle}},\ }in\
  \href@noop {} {\emph {\bibinfo {booktitle} {Theoretical and Computational
  Chemistry}}}\ (\bibinfo {year} {2004})\ p.\ \bibinfo {pages}
  {723}\BibitemShut {NoStop}%
\bibitem [{\citenamefont {Haverkort}(2016)}]{Haverkort:2016hz}%
  \BibitemOpen
  \bibfield  {author} {\bibinfo {author} {\bibfnamefont {M.~W.}\ \bibnamefont
  {Haverkort}},\ }\href@noop {} {\bibfield  {journal} {\bibinfo  {journal} {J.
  Phys.: Conf. Ser.}\ }\textbf {\bibinfo {volume} {712}},\ \bibinfo {pages}
  {012001} (\bibinfo {year} {2016})}\BibitemShut {NoStop}%
\bibitem [{\citenamefont {Nozi{\`e}res}\ \emph {et~al.}(1969)\citenamefont
  {Nozi{\`e}res}, \citenamefont {Gavoret},\ and\ \citenamefont
  {Roulet}}]{Nozieres:1969fk}%
  \BibitemOpen
  \bibfield  {author} {\bibinfo {author} {\bibfnamefont {P.}~\bibnamefont
  {Nozi{\`e}res}}, \bibinfo {author} {\bibfnamefont {J.}~\bibnamefont
  {Gavoret}}, \ and\ \bibinfo {author} {\bibfnamefont {B.}~\bibnamefont
  {Roulet}},\ }\href@noop {} {\bibfield  {journal} {\bibinfo  {journal} {Phys.
  Rev.}\ }\textbf {\bibinfo {volume} {178}},\ \bibinfo {pages} {1084} (\bibinfo
  {year} {1969})}\BibitemShut {NoStop}%
\bibitem [{\citenamefont {Doniach}\ and\ \citenamefont
  {Sunjic}(1970)}]{Doniach:1970wr}%
  \BibitemOpen
  \bibfield  {author} {\bibinfo {author} {\bibfnamefont {S.}~\bibnamefont
  {Doniach}}\ and\ \bibinfo {author} {\bibfnamefont {M.}~\bibnamefont
  {Sunjic}},\ }\href@noop {} {\bibfield  {journal} {\bibinfo  {journal} {J.
  Phys. C: Solid State Phys.}\ }\textbf {\bibinfo {volume} {3}},\ \bibinfo
  {pages} {285} (\bibinfo {year} {1970})}\BibitemShut {NoStop}%
\bibitem [{\citenamefont {Cornaglia}\ and\ \citenamefont
  {Georges}(2007)}]{Cornaglia:2007kn}%
  \BibitemOpen
  \bibfield  {author} {\bibinfo {author} {\bibfnamefont {P.~S.}\ \bibnamefont
  {Cornaglia}}\ and\ \bibinfo {author} {\bibfnamefont {A.}~\bibnamefont
  {Georges}},\ }\href@noop {} {\bibfield  {journal} {\bibinfo  {journal} {Phys.
  Rev. B}\ }\textbf {\bibinfo {volume} {75}},\ \bibinfo {pages} {115112}
  (\bibinfo {year} {2007})}\BibitemShut {NoStop}%
\bibitem [{\citenamefont {Bambynek}\ \emph {et~al.}(1977)\citenamefont
  {Bambynek}, \citenamefont {Behrens}, \citenamefont {Chen},\ and\
  \citenamefont {Crasemann}}]{Bambynek:1977wz}%
  \BibitemOpen
  \bibfield  {author} {\bibinfo {author} {\bibfnamefont {W.}~\bibnamefont
  {Bambynek}}, \bibinfo {author} {\bibfnamefont {H.}~\bibnamefont {Behrens}},
  \bibinfo {author} {\bibfnamefont {M.~H.}\ \bibnamefont {Chen}}, \ and\
  \bibinfo {author} {\bibfnamefont {B.}~\bibnamefont {Crasemann}},\ }\href@noop
  {} {\bibfield  {journal} {\bibinfo  {journal} {Rev. Mod. Phys.}\ }\textbf {\bibinfo {volume} {49}},\ \bibinfo {pages} {77} (\bibinfo
  {year} {1977})}\BibitemShut {NoStop}%
\bibitem [{\citenamefont {Altland}\ and\ \citenamefont
  {Simons}(2006)}]{AltlandSimons}%
  \BibitemOpen
  \bibfield  {author} {\bibinfo {author} {\bibfnamefont {A.}\ \bibnamefont
  {Altland}}\ and\ \bibinfo {author} {\bibfnamefont {B.}~\bibnamefont
  {Simons}},\ }in\
  \href@noop {} {\emph {\bibinfo {booktitle} {Condensed Matter Field Theory}}}\ (\bibinfo {year} {2006})\ p.\ \bibinfo {pages}
  {372ff}\BibitemShut {NoStop}%
\end{thebibliography}

%

\end{document}